\documentclass[12pt, a4paper]{article}
\usepackage{epsf}
\usepackage{cite}
\usepackage{amsmath,amssymb}
\input{colordvi.tex}
\usepackage[usenames,dvipsnames]{color}
\usepackage{graphicx}
\usepackage{ascmac}
\usepackage{hyperref}

\begin{document}

\begin{titlepage}

\begin{center}
\hfill TU-1175\\
\hfill KEK-QUP-2022-0021
\vskip .5in

{\Large \bf
Dark Photon Dark Matter from Cosmic Strings and \\ \vspace{2.5mm} 
Gravitational Wave Background
}

\vskip .5in

{\large
Naoya Kitajima$^{(a,b)}$ and
Kazunori Nakayama$^{(b,c)}$
}

\vskip 0.5in

$^{(a)}${\em 
Frontier Research Institute for Interdisciplinary Sciences, Tohoku University, Sendai 980-8578, Japan
}

\vskip 0.2in

$^{(b)}${\em 
Department of Physics, Tohoku University, Sendai 980-8578, Japan
}

\vskip 0.2in

$^{(c)}${\em 
International Center for Quantum-field Measurement Systems for Studies of the Universe and Particles (QUP), KEK, 1-1 Oho, Tsukuba, Ibaraki 305-0801, Japan
}

\end{center}
\vskip .5in

\begin{abstract}

Dark photon dark matter may be produced by the cosmic strings in association with the dark U(1) gauge symmetry breaking. We perform three-dimensional lattice simulations 
of the Abelian-Higgs model and follow the evolution of cosmic strings. In particular, we simulate the case of (very) light vector boson and find 
that such vector bosons are efficiently produced by the collapse of small loops while the production is inefficient in the case of heavy vector boson.
We calculate the spectrum of the gravitational wave background produced by the cosmic string loops for the light vector boson case and find characteristic features in the spectrum, which can serve as a probe of the dark photon dark matter scenario. In particular, we find that the current ground-based detectors may be sensitive to such gravitational wave signals and also on-going/future pulsar timing observations give stringent constraint on the dark photon dark matter scenario.

\end{abstract}

\end{titlepage}

\tableofcontents

\section{Introduction}

Dark photon is one of the candidates of dark matter (DM) of the Universe~\cite{Fabbrichesi:2020wbt,Caputo:2021eaa}. The dark photon may have string-theoretic origins~\cite{Goodsell:2009xc,Cicoli:2011yh} and phenomenologically it is characterized by the mass and coupling through the kinetic mixing with the Standard Model photon.
There are several known mechanisms for the dark photon production processes in order for the dark photon to be DM: gravitational production~\cite{Graham:2015rva,Ema:2019yrd,Ahmed:2020fhc,Kolb:2020fwh,Sato:2022jya,Redi:2022zkt}, gravitational thermal scattering~\cite{Tang:2017hvq,Garny:2017kha}, production through axion couplings~\cite{Agrawal:2018vin,Co:2018lka,Bastero-Gil:2018uel}, through Higgs couplings~\cite{Dror:2018pdh,Nakayama:2021avl} and cosmic strings~\cite{Long:2019lwl}.\footnote{
	The misalignment production scenario~\cite{Nelson:2011sf,Arias:2012az} suffers from serious instability or excluded by cosmological observations without significant modification of the cosmological models~\cite{Nakayama:2019rhg,Nakayama:2020rka} See also Ref.~\cite{Nakai:2022dni} for an alternative scenario.
}
In this paper we focus on the production through the cosmic strings.

If the dark photon, $A_\mu$, is associated with the dark U(1) gauge symmetry, it is natural that the finite dark photon mass arises from the Higgs mechanism, where the dark Higgs field $\Phi$ obtains a vacuum expectation value (VEV) and the dark U(1) symmetry is spontaneously broken. If the symmetry breaking happens after inflation, cosmic string networks appear in the Universe as topological defects~\cite{Vilenkin:2000jqa}. Once formed, the cosmic strings radiate dark photons if the dark photon is sufficiently light and the dark photon abundance produced in this way can be consistent with the present DM abundance~\cite{Long:2019lwl}.

One of the main purposes of our study is to confirm that the cosmic strings efficiently emit dark photons with the use of three-dimensional lattice simulation. On this point, one should note that there are already a series of studies for the case of global strings, i.e., the axion emission from cosmic strings associated with the global U(1) symmetry breaking~\cite{Davis:1985pt,Davis:1986xc,Vilenkin:1986ku,Harari:1987ht,Davis:1989nj,Dabholkar:1989ju,Hagmann:1990mj,Battye:1993jv,Battye:1994au,Yamaguchi:1998gx,Yamaguchi:1999yp,Yamaguchi:1999dy,Hagmann:2000ja,Hiramatsu:2010yu,Hiramatsu:2012gg,Fleury:2015aca,Klaer:2017qhr,Gorghetto:2018myk,Kawasaki:2018bzv,Buschmann:2019icd,Hindmarsh:2019csc,Klaer:2019fxc,Gorghetto:2020qws,Hindmarsh:2021vih,Buschmann:2021sdq,Blanco-Pillado:2022axf}.
An important property of the global strings is that the average string number per Hubble volume grows logarithmically with time:
$\xi(t) = \beta + \alpha\log(m_r/H)$, which we call the logarithmic violation of the scaling law~\cite{Gorghetto:2018myk,Kawasaki:2018bzv,Buschmann:2019icd,Klaer:2019fxc,Gorghetto:2020qws,Buschmann:2021sdq}.
On the other hand, Refs.~\cite{Hindmarsh:2019csc,Hindmarsh:2021vih} claim the conventional scaling law: $\xi(t) = {\rm const} \simeq 1$.
Although these results have been obtained for global strings, one may expect that the production of the longitudinal vector boson should be similar to the axion thanks to the Goldstone boson equivalence theorem if the momentum is high enough. We will reveal it by an explicit calculation.

Another main purpose of this paper is to estimate the gravitational wave (GW) background spectrum in the cosmic string scenario for the dark photon DM. The estimation of GW background from the cosmic strings, in particular from cosmic string loops, have been done in many works mostly in the case of local strings~\cite{Vachaspati:1984gt,Garfinkle:1987yw,Caldwell:1991jj,Caldwell:1996en,Damour:2000wa,Damour:2001bk,Damour:2004kw,Siemens:2006yp,DePies:2007bm,Olmez:2010bi,Binetruy:2012ze,Kuroyanagi:2012wm,Kuroyanagi:2012jf,Ringeval:2017eww,Cui:2017ufi,Gouttenoire:2019kij}.
Recently GWs from global string have been calculated in connection with the recent development of the global string simulation with the improved scaling properties~\cite{Gorghetto:2021fsn,Chang:2021afa}.
The inclusion of axion or vector boson (dark photon) emission drastically affects the resulting GW spectrum, since the most energy of string loops goes to the axion or dark photon emission while only tiny fraction of the loop energy is consumed as GW emission.\footnote{
	The cusps on the string loops may also emit high-energy particles~\cite{Blanco-Pillado:1998tyu,Olum:1998ag}. See also Refs.~\cite{Hindmarsh:2017qff,Matsunami:2019fss,Saurabh:2020pqe} where efficient dark photon/Higgs emission from local-type strings has also been reported. 
 Phenomenology/cosmology of cosmic strings due to heavy particle radiation have been discussed in Refs.~\cite{Jeannerot:1999yn,Cui:2007js,Cui:2008bd,Kawasaki:2011dp,Miyamoto:2012ck,Hindmarsh:2022awe}. 
}
Even more complexity arises when the dark photon is light but has finite mass. In such a case, loops longer than some threshold cannot emit dark photon while shorter ones can emit it.
Thus the cosmic strings in this scenario will be a mixture of “local-type” and “global-type” strings.
Such nontrivial structures should be imprinted in the shape of the GW spectrum. Ref.~\cite{Long:2019lwl} already pointed out such a possibility, but detailed calculations have not been performed. We show explicitly the GW spectrum as a probe of the dark photon DM scenario.

This paper is organized as follows. 
In Sec.~\ref{sec:sim} we perform a three-dimensional lattice simulation of the local strings with light dark photon. 
In Sec.~\ref{sec:GW} we estimate the GW background from the cosmic string loops in the dark photon DM scenario.
In Sec.~\ref{sec:con} we give conclusions and discussion.

\section{Simulation of Abelian-Higgs string networks} \label{sec:sim}

In this section, we numerically show the evolution of Abelian-Higgs string networks with both local-type and near-global-type cases.
First, we focus on the scaling behavior for the cosmic string length. Then, we show the emission of dark photons from the string network.

\subsection{Model}

Let us consider the Abelian-Higgs model in the hidden sector with the dark U(1) gauge symmetry given by the following Lagrangian,
\begin{align}
\mathcal{L} = (D_\mu \Phi)^* D^\mu \Phi -\frac{1}{4} F_{\mu \nu}F^{\mu \nu}-\frac{\lambda}{4}(|\Phi|^2-v^2)^2,
\end{align}
where $\Phi$ is the dark Higgs field, $D_\mu = \partial_\mu-ie A_\mu$ is the covariant derivative operator with $A_\mu$ being the dark photon field, $F_{\mu\nu} = \partial_\mu A_\nu-\partial_\nu A_\mu$ is the field strength tensor, $e$, $\lambda$ and $v$ are respectively the gauge coupling constant, the scalar self coupling constant and the vacuum expectation value.
Since $|\Phi|=v$ at the minimum of the potential, the U(1) symmetry is spontaneously broken and cosmic strings are formed in association with the symmetry braking.
The equations of motion are derived as follows,
\begin{align}
&\frac{1}{\sqrt{-g}} D_\mu(\sqrt{-g}D^\mu \Phi) + \frac{\partial V}{\partial \Phi^*} = 0,\\
&\frac{1}{\sqrt{-g}} \partial_\mu(\sqrt{-g} F^{\mu\nu}) -2e\,{\rm Im} (\Phi^* D^\nu \Phi) = 0.
\end{align}
Here and in what follows, we assume the flat Friedmann-Lemaitre-Robertson-Walker (FLRW) metric as the background spacetime and use the conformal time, $\tau$, as time variable. In addition, we impose the temporal gauge, $A_0=0$, throughout this paper. The system of evolution equations can be written as
\begin{align}
&\Phi''+2\mathcal{H}\Phi'-D_i D_i \Phi + a^2 \frac{\partial V}{\partial \Phi^*} = 0, \label{eq:phi-eom}\\ 
&E_i'+\partial_j F_{ij} -2e a^2\,{\rm Im}(\Phi^*D_i \Phi) = 0,\label{eq:A-eom}\\
&\partial_i E_i -2e a^2\,{\rm Im}(\Phi^*\Phi') = 0, \label{eq:GaussLaw}
\end{align}
where the prime denotes the derivative with respect to the conformal time, $a$ is the scale factor, $\mathcal{H}=a'/a$, $E_i \equiv F_{0i}$ and repeated indices are summed.\footnote{The electric field $E_i$ is defined through the derivative with respect to the conformal time $\tau$ in this paper.}
Note that two of these equations (Eq.~(\ref{eq:phi-eom}) and (\ref{eq:A-eom})) determine the field dynamics and the Eq.~(\ref{eq:GaussLaw}) is the constraint equation corresponding to the Gauss's law.
Here and in what follows, we assume the radiation-dominated universe, i.e. $a \propto \tau$ and $\mathcal{H} = 1/\tau$ and the energy density of the Abelian-Higgs system is negligible with respect to the background radiation density.
Note that the dynamics depends only on the ratio $\lambda/(2e^2)$ after redefining the field values. It corresponds to the square of the ratio between the mass of the radial component of the scalar field and the mass of the dark photon $m_A = \sqrt{2}ev$ around the symmetry-breaking vacuum. Specifically, the dark photon is much lighter than the radial mode for $e^2 \ll \lambda$ and one can expect that string configuration approaches to the global string one. In the following, we numerically show the scaling behavior and emission of dark photons in both local and near-global cases. In the following numerical analysis, we set $\lambda=2$ throughout.

\subsection{Numerical simulation}

One can follow the evolution of Abelian-Higgs string network by solving the system of equations (\ref{eq:phi-eom}),(\ref{eq:A-eom}) under the constraint equation (\ref{eq:GaussLaw}). In general, the dynamics of topological defects is highly nonlinear and hence we adopt three-dimensional lattice simulation.

The Abelian-Higgs lattice simulation has been performed in the literature \cite{Vincent:1997cx,Moore:2001px,Bevis:2006mj,Hindmarsh:2008dw,Dufaux:2010cf,Hiramatsu:2013tga,Daverio:2015nva,Hindmarsh:2017qff,Correia:2018gew,Correia:2020yqg,Blanco-Pillado:2023sap} and we follow the formulation based on the lattice gauge theory \cite{Bevis:2006mj}. In this formalism, the scalar field is placed at grid points and the gauge field is placed at links between two grid points. Discretized covariant derivative is defined so as to satisfy the exact gauge invariance (up to the machine precision) while it coincides with the continuous one within the second order accuracy with respect to the lattice spacing.
Field values are updated following Eqs. (\ref{eq:phi-eom}),(\ref{eq:A-eom}) by using the Leap-frog method.
The Gauss's law (\ref{eq:GaussLaw}) is also discretized and, in this formalism, it is exactly satisfied at each grid point at each step.\footnote{Practically, the left-hand-side in (\ref{eq:GaussLaw}) cannot be exactly zero in the simulation due to the finite precision of the floating point number on the computer. By changing the precision of data, we have explicitly checked that the residual error is just due to the cancellation of significant digits.}

As the initial condition, we set the random value for the phase component of the complex scalar field with the uniform probability density but the radial component is fixed to be $|\Phi|=v$ everywhere. 
Initial random fields are generated in the Fourier space with the cutoff scale $k_c/a=H$, above which we set zero for the field value.
The time derivative of the scalar field and each component of the gauge field is initially set to be zero. Regarding the identification of cosmic string in the lattice space, we follow the method in  \cite{Kajantie:1998bg} using the gauge-invariant sum of phase differences.

In order to evaluate the scaling behavior of the network of cosmic strings, we introduce two quantities. One is a mean separation between neighboring cosmic strings which is conventionally adopted in the context of the Abelian-Higgs string simulation. It has a dimension of length and  defined by
\begin{align}
d_{\rm sep} = \sqrt{\frac{V_{\rm box}}{\ell_{\rm tot}}},
\end{align}
where $V_{\rm box}$ is the physical volume of the simulation box and $\ell_{\rm tot}$ is the total physical length of cosmic strings.
When the string network follows the scaling regime, one can express it as $m_r d_{\rm sep}  = a m_r\tau + b$ with $a$ and $b$ being some numerical constants and $m_r \equiv \sqrt{\lambda/2}v$.
The other is defined by the ratio of the string length to the Hubble length in one Hubble volume of the universe, which is conventionally used in the axion string simulation. It is dimensionless quantity defined by
\begin{align}
\xi = \frac{\ell_{\rm tot} t^2}{V_{\rm box}},
\end{align}
with $t$ the cosmic time.
In this case, $\xi=\mathcal O(1)$ (constant) when the string network follows the scaling regime.

\begin{figure}
\begin{center}
   \includegraphics[width=8cm]{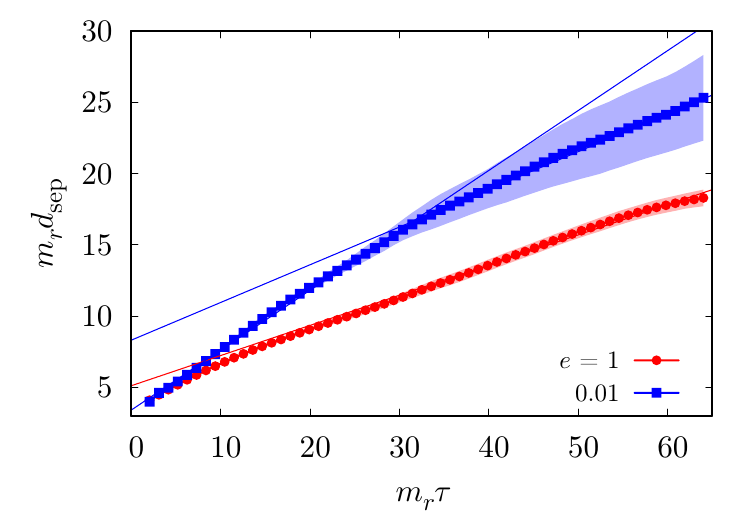}
    \includegraphics[width=8cm]{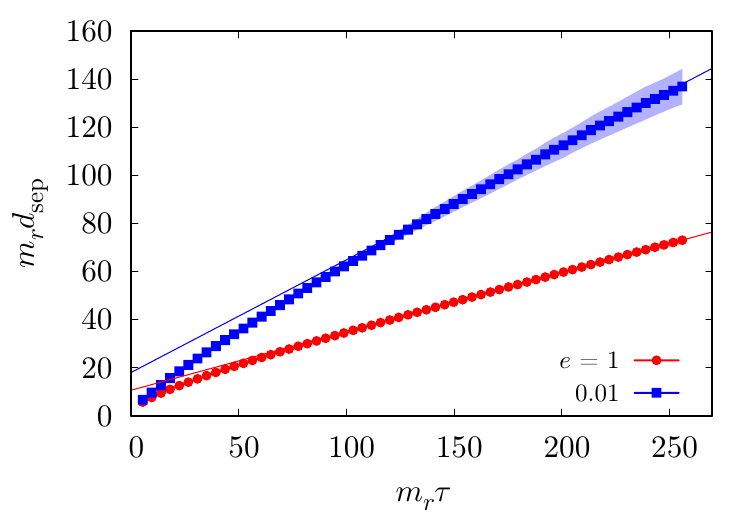}
  \end{center}
  \caption{Time evolution of the mean separation of cosmic strings for physical (left) and fat (right) string cases. The red and blue lines correspond to $e=1$ and $e=0.01$ respectively. Shaded region shows the 1-$\sigma$ error. Data points are fitted by linear functions shown by thin dashed lines.}
  \label{fig:dsep}
\end{figure}

\begin{figure}
\begin{center}
   \includegraphics[width=8cm]{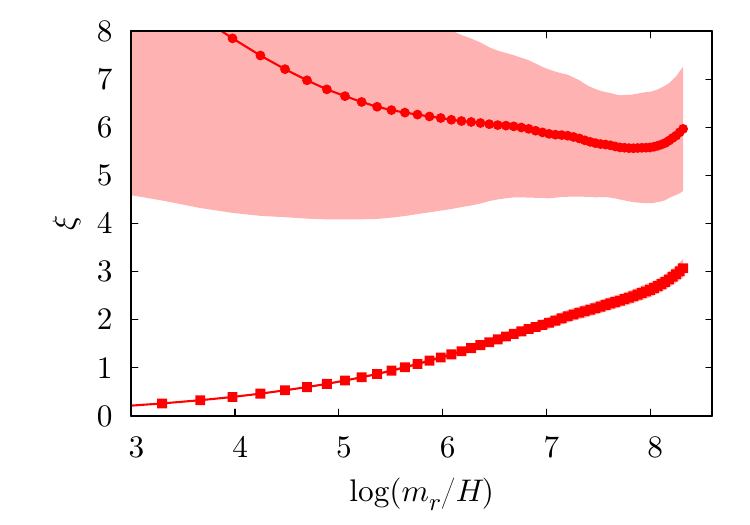}
    \includegraphics[width=8cm]{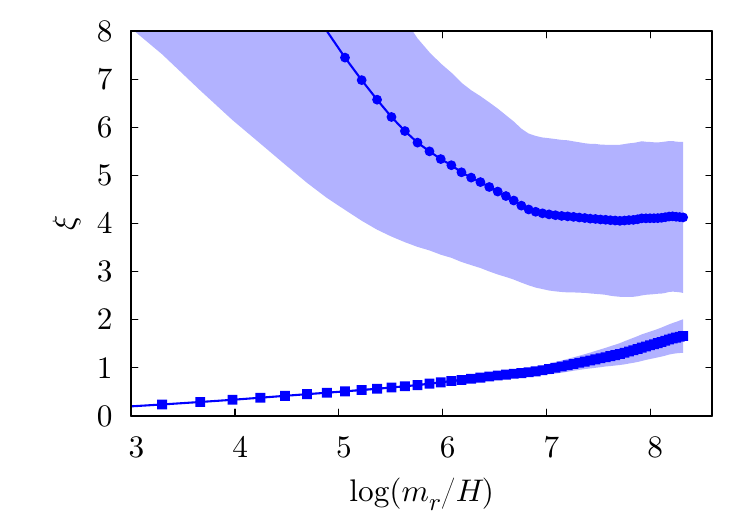}
  \end{center}
  \caption{Square points (lower points) show time evolution of $\xi$ for the physical strings with $e=1$ (left) and $e=0.01$ (right). Circle points (upper points) correspond to the corrected value $\hat\xi$. Shaded region shows the 1-$\sigma$ error.}
  \label{fig:xi_phys}
\end{figure}

\begin{figure}
\begin{center}
   \includegraphics[width=8cm]{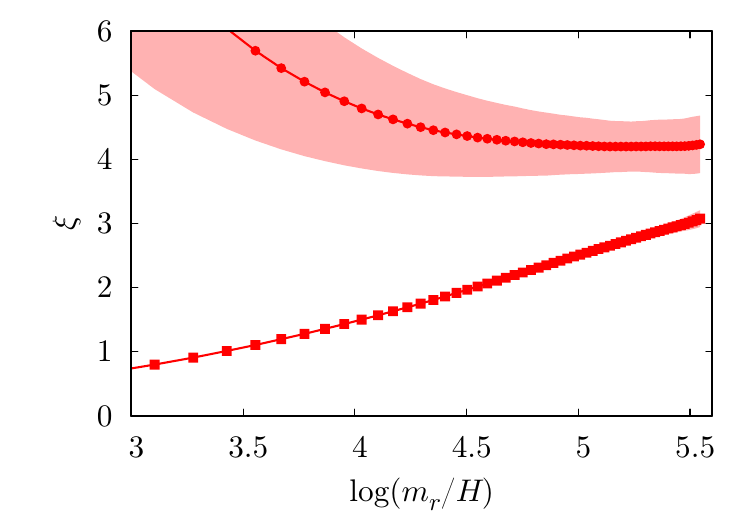}
    \includegraphics[width=8cm]{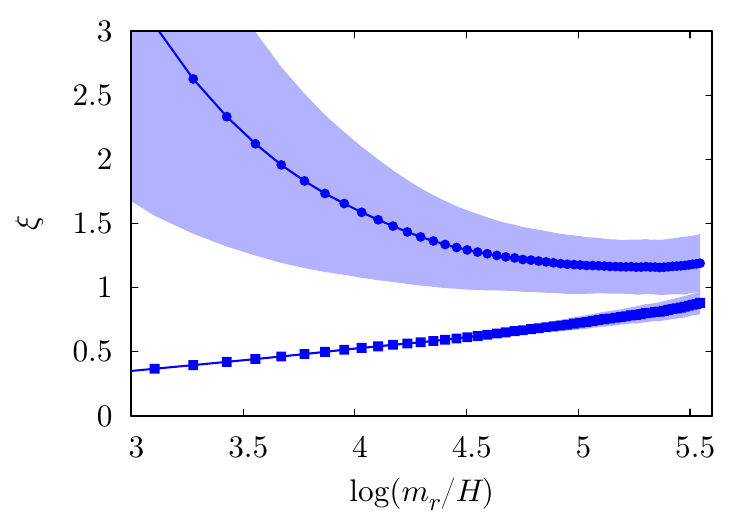}
  \end{center}
  \caption{Same as Fig.~\ref{fig:xi_phys} but for the fat string case.}
  \label{fig:xi_fat}
\end{figure}

\begin{figure}
\begin{center}
   \includegraphics[width=8cm]{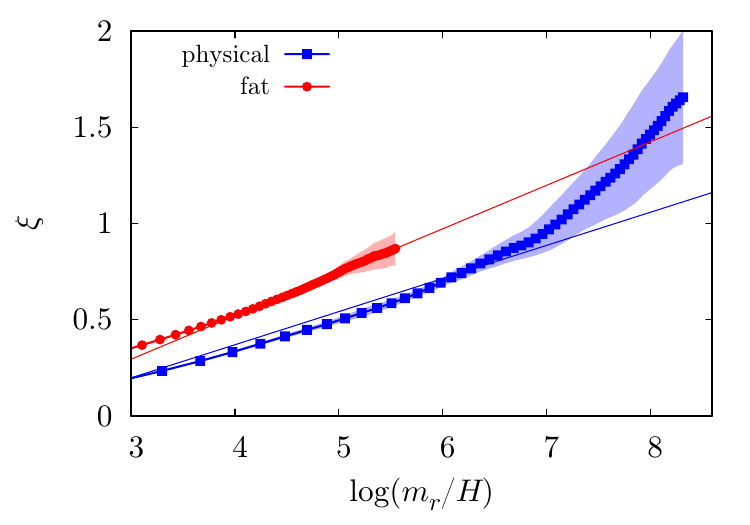}
  \end{center}
  \caption{Time evolution of the string length parameter $\xi$ for physical and fat string cases shown by the blue square and red circle points respectively. Shaded region shows the 1-$\sigma$ error. Data points are fitted by log functions, $\alpha \log(m_r/H)+\beta$, shown by thin lines.}
  \label{fig:xi_log}
\end{figure}

\begin{table}
\begin{center}
\begin{tabular}{c c c c c c c c}
\hline
~& grid & $m_r L$ & $m_r \Delta x$ & $e$ & $m_r \tau$ & $a$ & $b$ \\
\hline
physical string & $4096^3$ & 64 & 1/64 & 1 & 33.6\,-\,64.0 & $0.21 \pm 0.0043$ & $5.1 \pm 0.21$  \\
physical string & $4096^3$ & 64 & 1/64 & 0.01 & 2.05\,-\,32.5 & $0.42 \pm 0.0038$ & $3.4 \pm 0.073$  \\
physical string & $4096^3$ & 64 & 1/64 & 0.01 &33.6\,-\,64.0 & $0.27 \pm 0.019$ & $8.4 \pm 0.96$  \\
fat string & $1024^3$ & 512 & 1/2 & 1 & 133\,-\,256 & $0.24 \pm 0.0013$ &  $11 \pm 0.25$\\
fat string & $1024^3$ & 512 & 1/2 & 0.01 & 133\,-\,256 & $0.47 \pm 0.0081$ &  $18 \pm 1.6$\\
\hline
\end{tabular}
\end{center}
\caption{Simulation setup and linear fitting parameters of the mean string separation in terms of the conformal time, defined by $m_r d_{\rm sep} = a m_r \tau + b$.}
\label{tab:fitting}
\end{table}

\begin{table}
\begin{center}
\begin{tabular}{c c c c c c c c}
\hline
~& grid & $m_r L$ & $m_r \Delta x$ & $e$ & $m_r \tau$ & $\alpha$ & $\beta$ \\
\hline
physical string & $4096^3$ & 64 & 1/64 & 0.01 & 2.05\,-\,32.5 & $0.17 \pm 0.0034$ & $-0.32 \pm 0.019$  \\
fat string & $1024^3$ & 512 & 1/2 & 0.01 & 133\,-\,256 & $0.23 \pm 0.018$ &  $-0.38 \pm 0.096$\\
\hline
\end{tabular}
\end{center}
\caption{Log fitting parameters of the string length parameter, defined by $\xi = \alpha \log(m_r/H)+ \beta$.}
\label{tab:log_fitting}
\end{table}

We start the simulation with the initial conformal time $m_r \tau = 1$ and end it at $a \Delta x = m_r^{-1}$ and $aL = 2H^{-1}$ for physical and fat string cases respectively (see below), with $\Delta x$ and $L$ being respectively the comoving lattice spacing and the comoving box size. The Setup of our simulation is summarized in Table \ref{tab:fitting} and we start the simulation with $m_r\tau = 1$.\footnote{We have used supercomputer ``AOBA'' at Cyberscience Center, Tohoku University.}
Note that, in addition to the ``physical'' string case with $\lambda$ and $e$ being constant with time, we also consider the ``fat'' string case where $\lambda$ and $e$ are replaced by the time-dependent quantities, $\lambda a(t)^{-2}$ and $e a(t)^{-1}$ respectively, in order to keep the string core size and the vector core size larger than the spatial resolution of the lattice simulation. We have run 5 (10) simulations for the physical (fat) string case.
Fig.~\ref{fig:dsep} shows the time evolution of the mean string separation, $d_{\rm sep}$, for physical (left) and fat (right) string cases. Linear fitting lines are overlaid in these figures.
The latter half of data points ($m_r\tau = 33.6$-64.0 and 133-256 for physical and fat string case respectively) are used for the linear fitting lines, $a m_r\tau+b$ with $a$ and $b$ being fitting parameters.
In addition, we also use the former half of data points for the physical string case with $e=0.01$.
The linear fitting parameters obtained by our simulations are summarized in Table \ref{tab:fitting}. The obtained values are consistent with previous studies, \cite{Daverio:2015nva} for local string and \cite{Hindmarsh:2019csc} for global string, corresponding respectively to $e=1$ and $0.01$ in our case. 
One can clearly see that in the physical string case with $e=0.01$, there is a transition around $m_r \tau = 30$. This implies that the string with $e=0.01$ can be regarded as the near-global string in the early time and it finally behaves like the local string.
%

Square points (lower points) in Fig.~\ref{fig:xi_phys} and \ref{fig:xi_fat} show the time evolution of $\xi$ for the physical and fat string cases respectively. Because of the constant shift from the linear function in $d_{\rm sep}$ ($b$ in the above expression), $\xi$ cannot settle down to a constant value even in the local string case, at least in our limited simulation time.  
The constant shift can be a numerical artifact depending on the initial condition. Then, let us remove it following the analysis in \cite{Hindmarsh:2019csc}.
Here the string length density parameter is replaced by 
\begin{align}
\hat\xi = \frac{\ell_{\rm tot} (t+t_0)^2}{V_{\rm box}} = \frac{\tilde{\ell}_{\rm tot} (\tau+\tau_0)^2}{4\tilde{V}_{\rm box}},
\end{align}
where $t_0$ and $\tau_0$ represent the constant shift with respect to the cosmic time and conformal time respectively and the tilde denotes the comoving length/volume. $\tau_0$ can be computed by the linear fitting of $d_{\rm sep}$ (namely, $\tau_0 = b/a$). The corrected value $\hat\xi$ is shown by the circle points (upper points) in Fig.~\ref{fig:xi_phys} and \ref{fig:xi_fat}. One can see that the corrected value $\hat\xi$ approaches a constant value while the original value $\xi$ increases with time.\footnote{In the physical string case with $e=1$, sudden increase near the end of the simulation is not physical but it is due to the finite box size effect.}
%
%
%
We also fitted the string length parameter $\xi$ by the log function, $\alpha \log(m_r/H)+\beta$ in both physical and fat string cases with $e=0.01$. We adopted the former half of the data for physical string case because the late-time dynamics behaves like a local string one. The data can be well fitted by the log function and the fitting parameters are summarized in Table \ref{tab:log_fitting}. Those parameters are near the values reported in \cite{Gorghetto:2018myk,Kawasaki:2018bzv,Gorghetto:2020qws} for the axion (global) string case. 
After the Hubble parameter becomes smaller than the dark photon mass, the dark photon emission from the loop collapse is kinematically suppressed. Then, one can expect that, even in the case with $e \ll 1$, the cosmic strings behave like local strings for $m_A > H$ and the scaling violation is no longer sustained. Thus, one can expect that $\xi$ tends to be constant after that. 
However, it is still controversial whether the scaling violation indeed occurs or not. In particular, the error becomes large near the end of simulation, which makes it difficult to conclude which scenario is true. Thus, one needs larger-scale (longer-time) simulation with smaller gauge coupling constant to confirm whether the scaling violation occurs or not. We leave it for future work and in what follows we take $\xi$ as a free parameter, which takes $O(1)$\,-\,$O(10)$.

\begin{figure}[t]
\begin{center}
   \includegraphics[width=5.4cm]{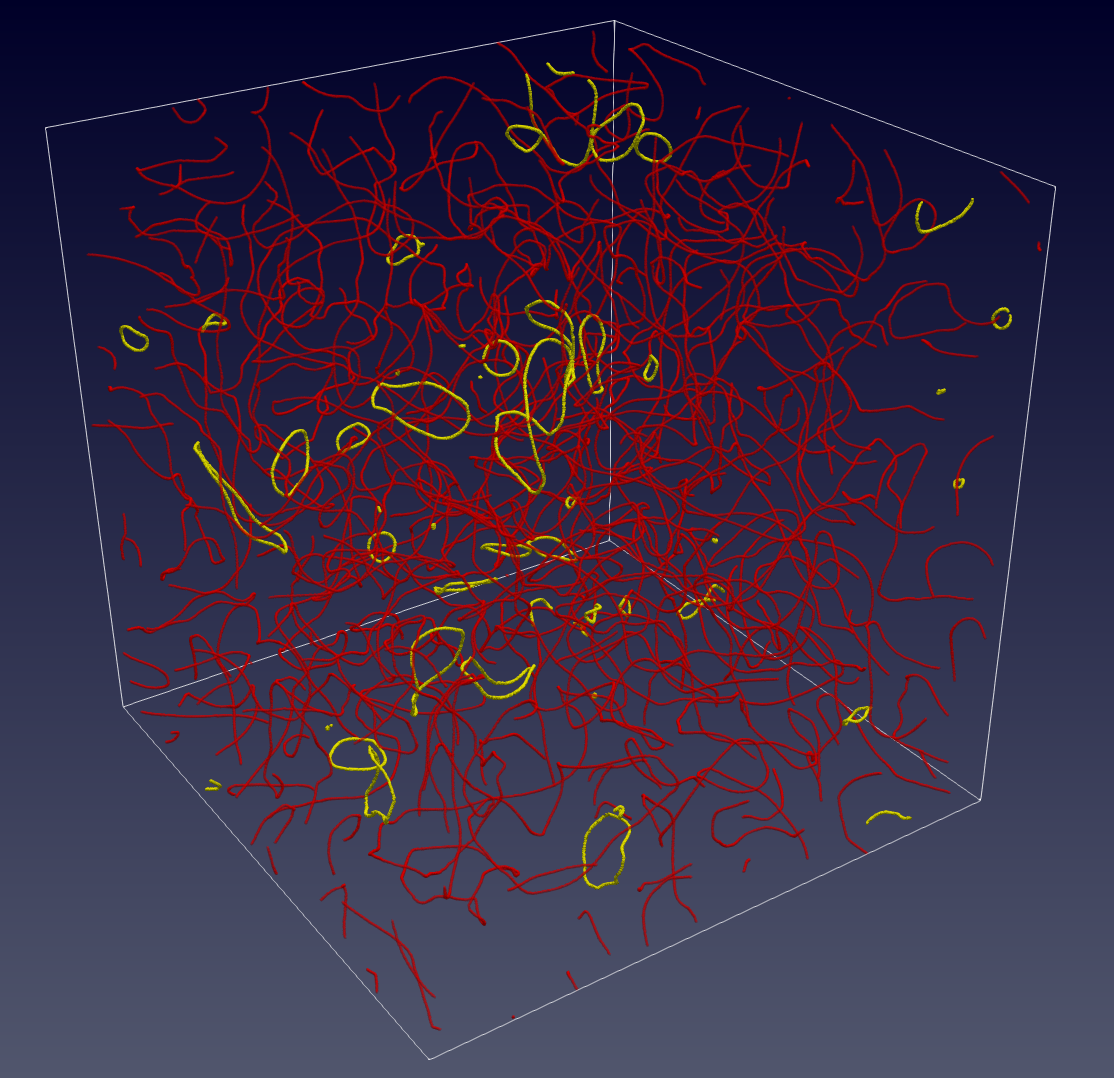}
   \includegraphics[width=5.4cm]{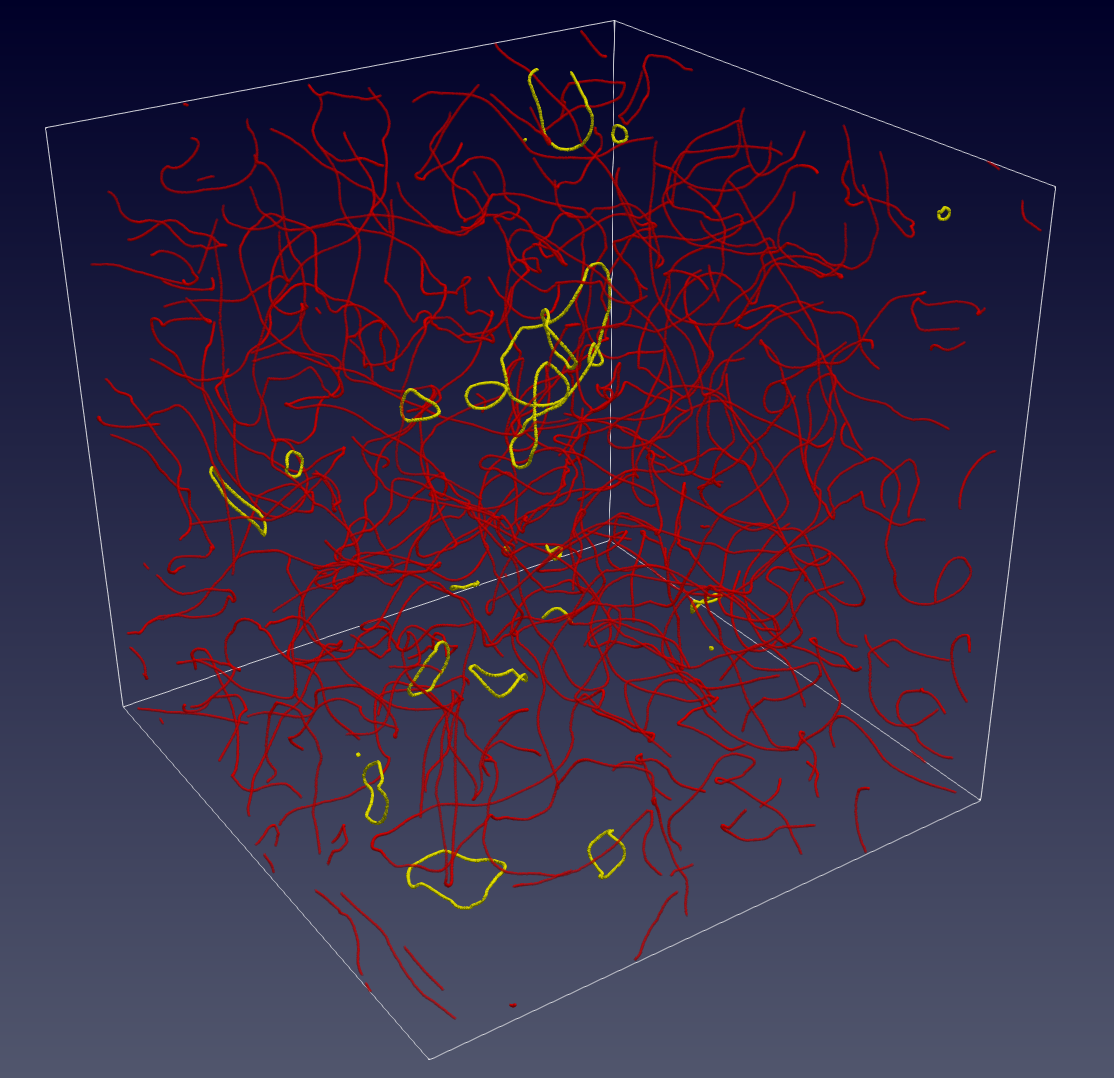}
   \includegraphics[width=5.4cm]{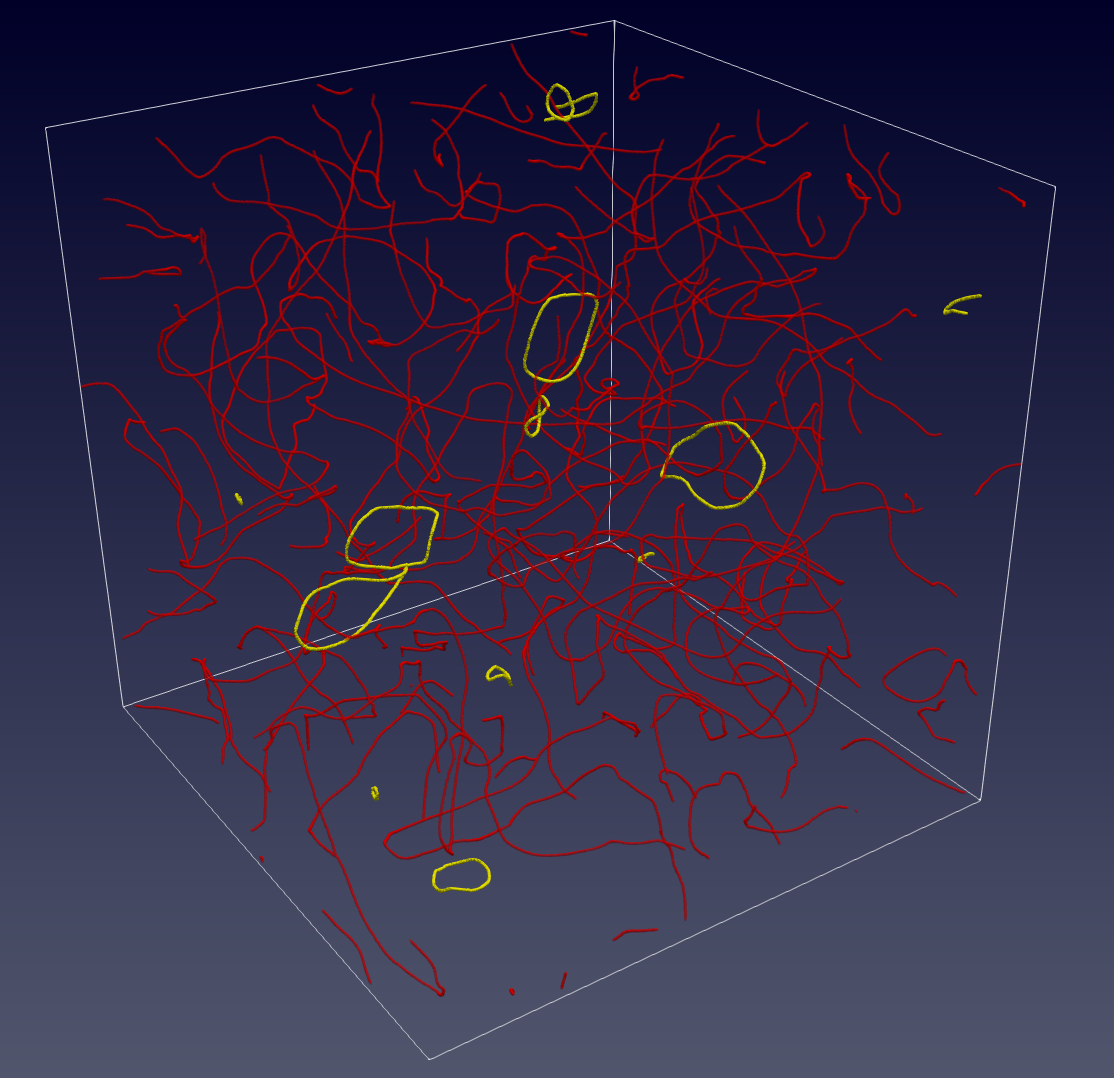}
   \end{center}
   \vspace{-6mm}
  \caption{Snapshots of the simulation for $\lambda=2$, $e=1$ and $\log(m_r/H)=3.4$, $4.6$, $5.3$ from left to right. Loops with comoving radius smaller than $L_{\rm box}/(2\pi)$ are highlighted.}
  \label{fig:snapshot1}
\end{figure}
\begin{figure}[t]
\begin{center}
   \includegraphics[width=5.4cm]{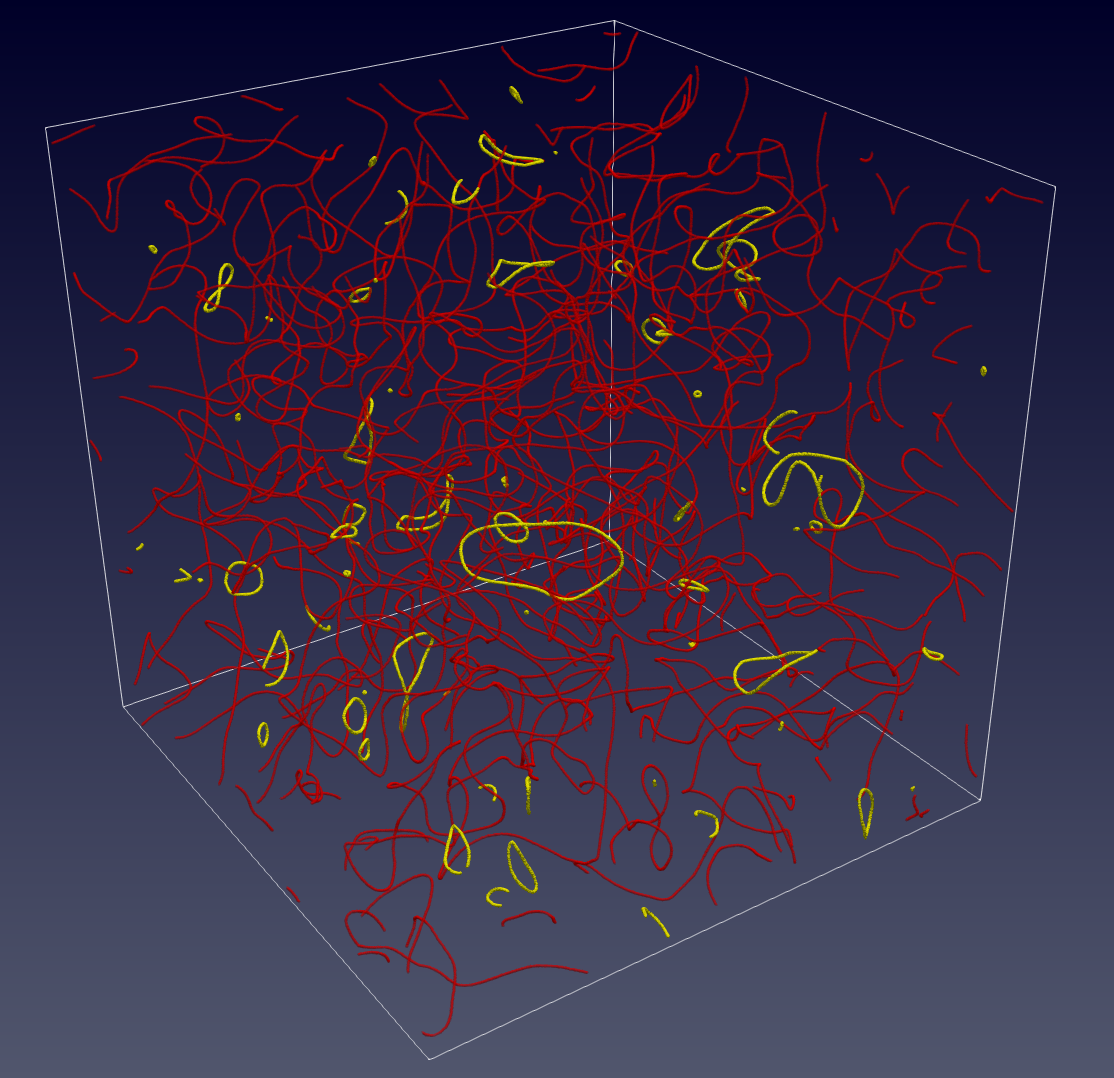}
   \includegraphics[width=5.4cm]{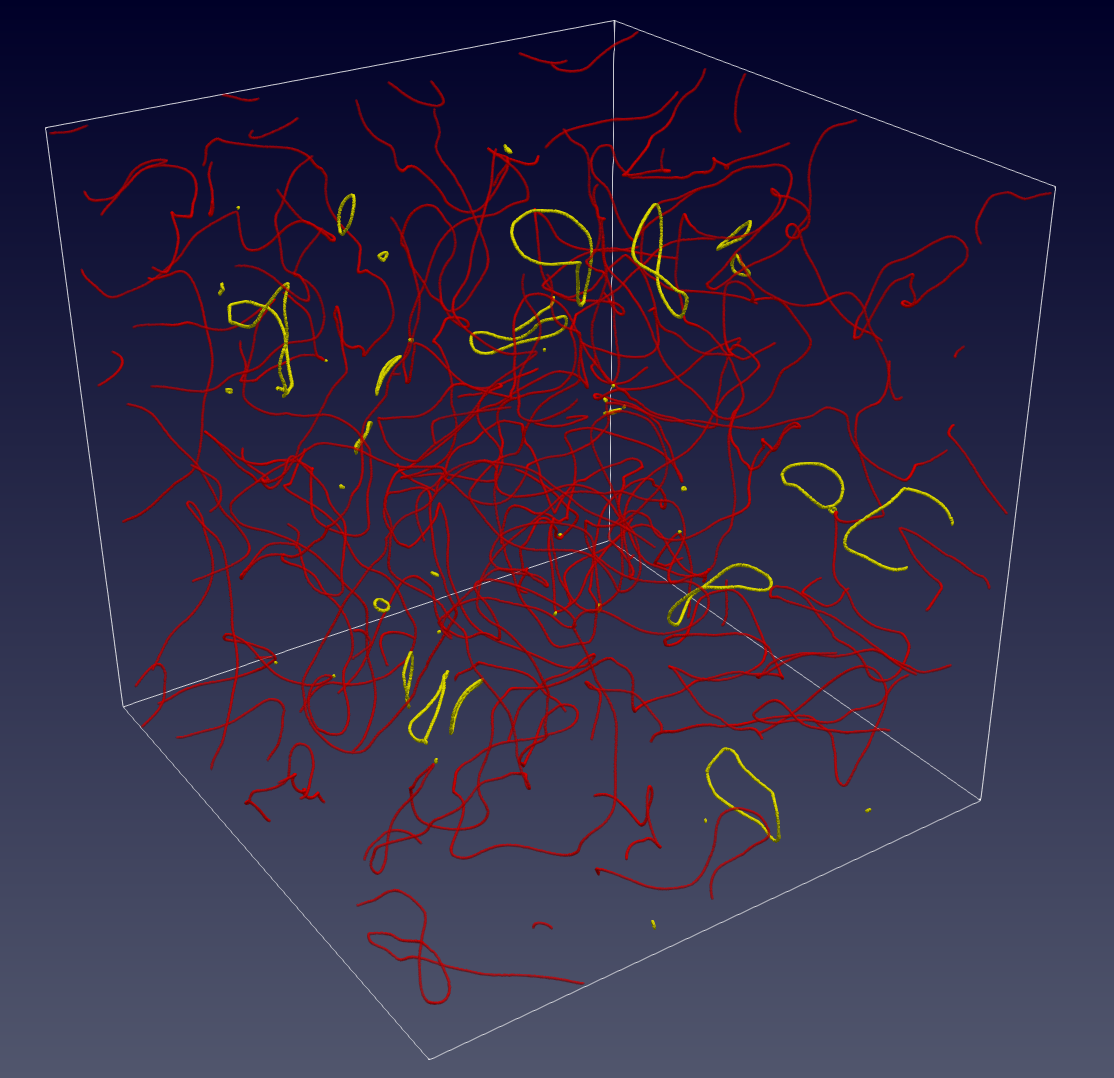}
   \includegraphics[width=5.4cm]{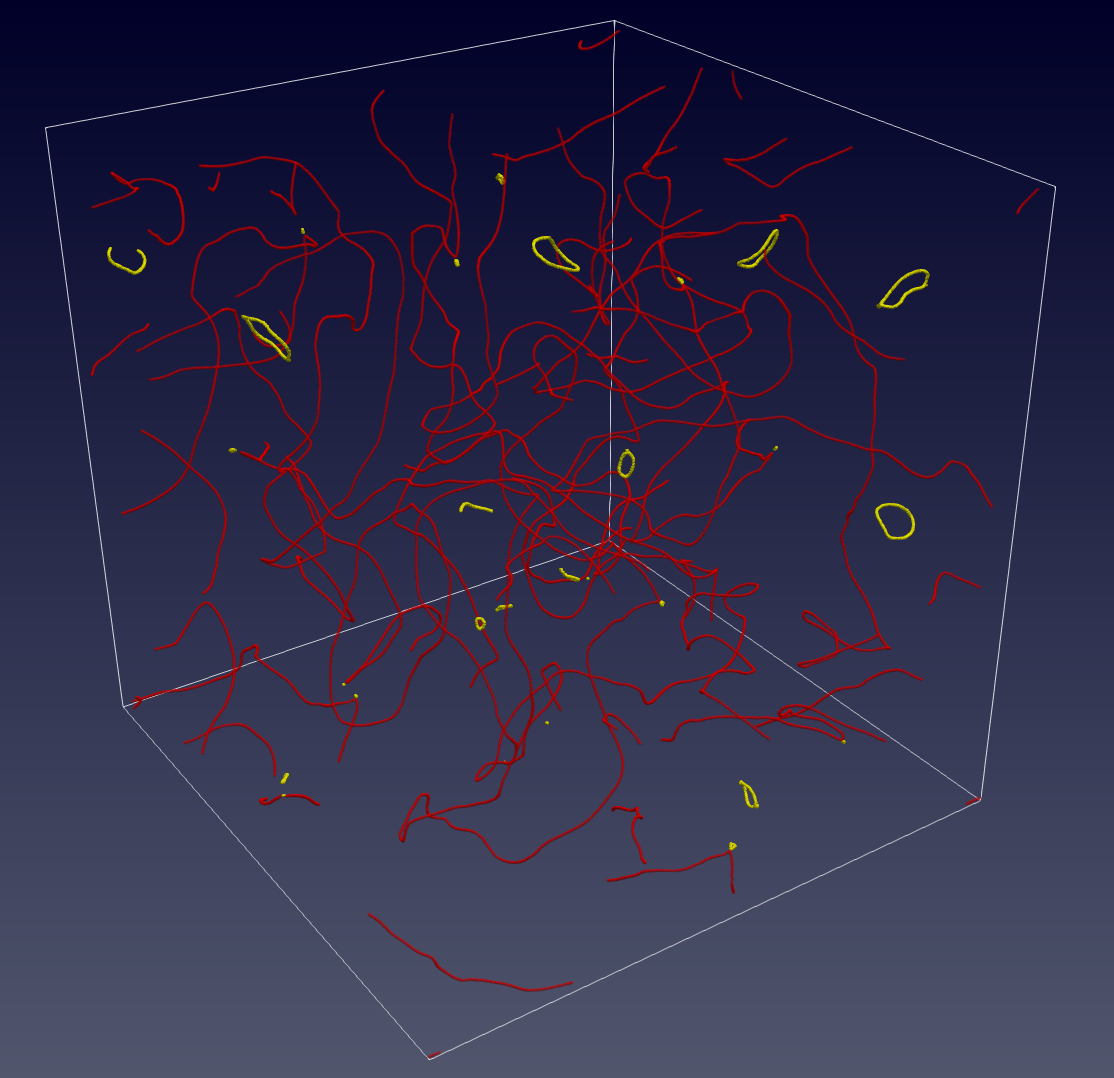}
   \end{center}
   \vspace{-6mm}
  \caption{Same as Fig~\ref{fig:snapshot1} but $e=0.01$.}
  \label{fig:snapshot2}
\end{figure}

In the rest of this paper, we consider only the physical string.
Figs.~\ref{fig:snapshot1} and \ref{fig:snapshot2} show snapshots of time slices with $\log(m_r/H) = 3.4$, $4.6$, $5.3$ from left to left to right for $e=1$ and $0.01$ respectively. One can see more abundant small loops in the case with $e=0.01$ corresponding to the near-global string case especially in the late time. It is due to the efficient vector boson emission as shown below.

\begin{figure}
\begin{center}
   \includegraphics[width=5.4cm]{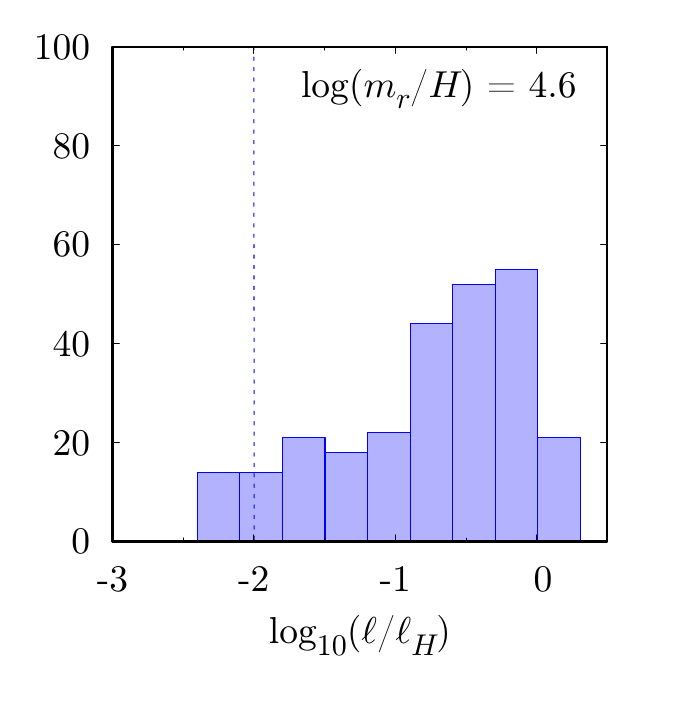}
    \includegraphics[width=5.4cm]{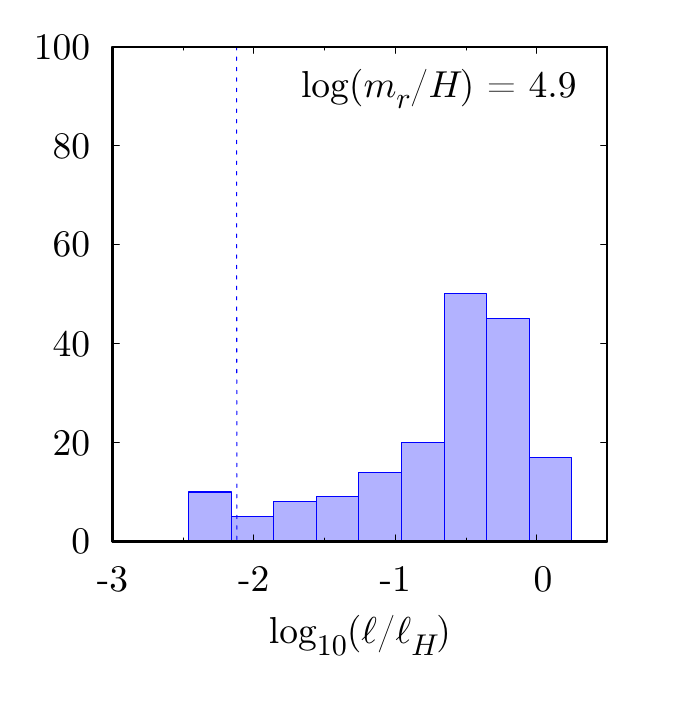}
    \includegraphics[width=5.4cm]{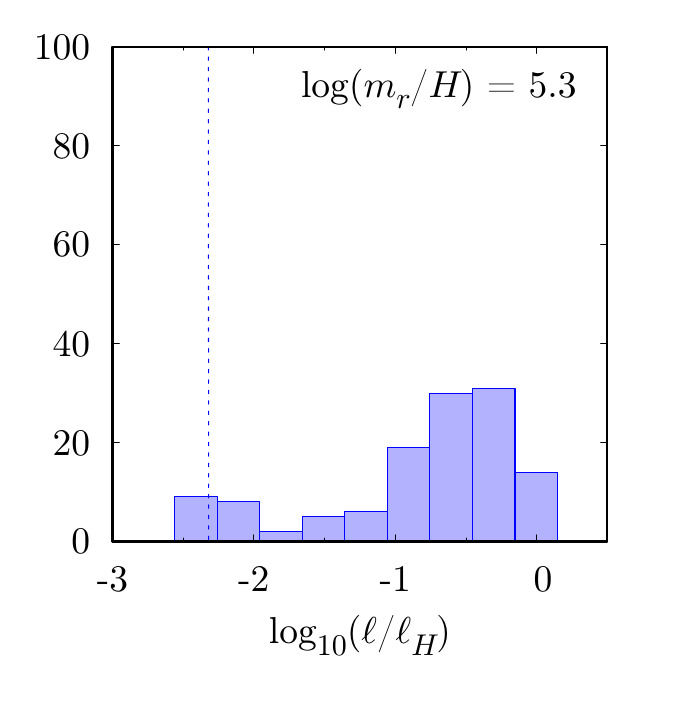}
    \includegraphics[width=5.4cm]{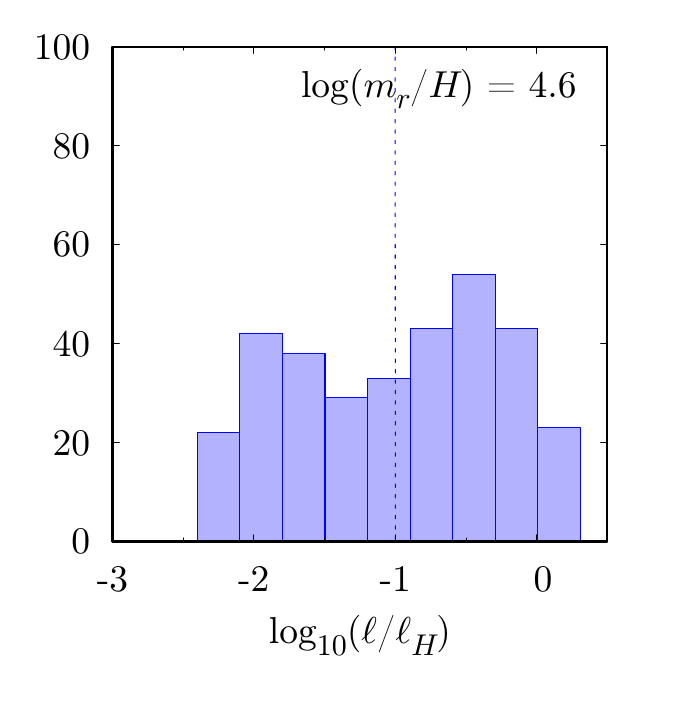}
     \includegraphics[width=5.4cm]{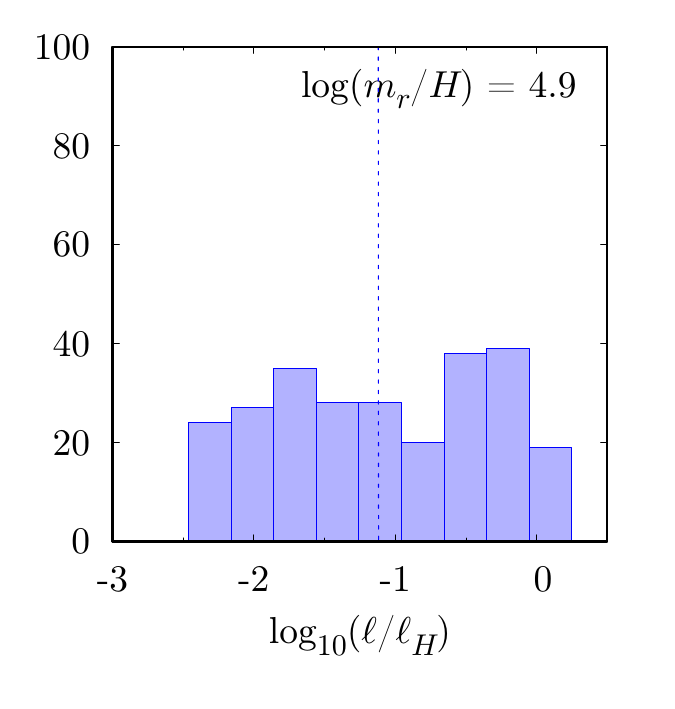}
    \includegraphics[width=5.4cm]{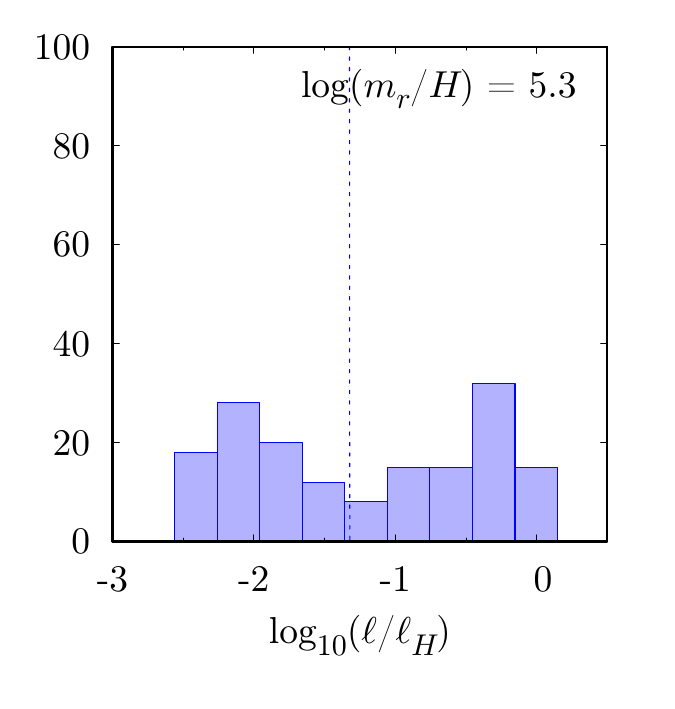}
    \includegraphics[width=5.4cm]{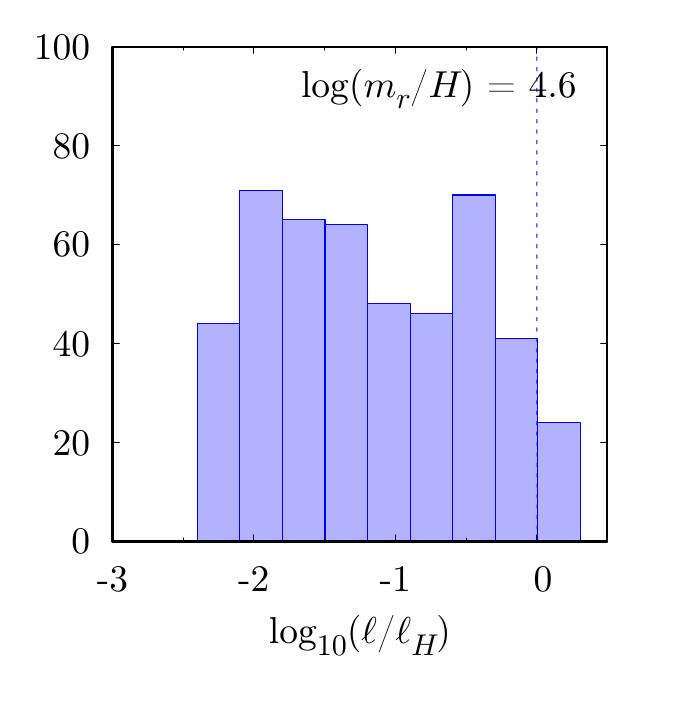}
     \includegraphics[width=5.4cm]{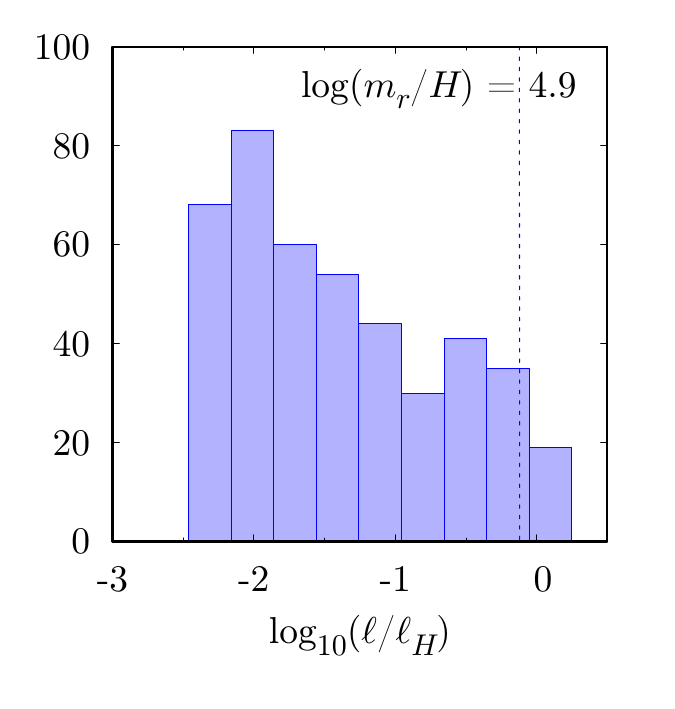}
    \includegraphics[width=5.4cm]{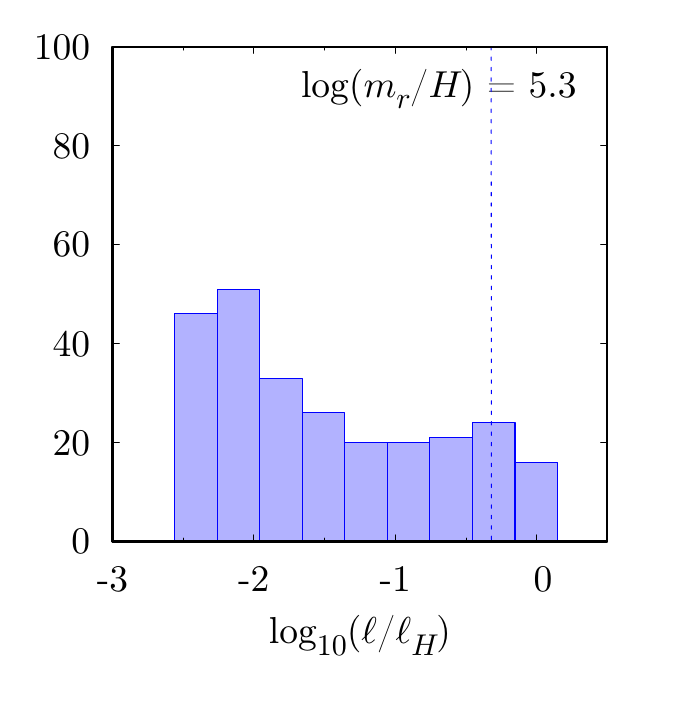}
  \end{center}
  \caption{The number distribution of loops with respect to the loop size for $e=1$ (top), $0.1$ (middle), $0.01$ (bottom). In the horizontal axis, $\ell_H \equiv H^{-1}$ is the Hubble length. Time slices correspond to $\log(m_r/H)=4.6$, $4.9$, $5.3$ from left to right. The vertical dashed line corresponds to $m_A \ell =1$.}
  \label{fig:loop_dist}
\end{figure}

The loop production efficiency is quantitatively shown in Figure \ref{fig:loop_dist}, showing the number distribution of loops with logarithmic interval of loop size ($\ell$) over the Hubble length ($\ell_H = H^{-1}$). Those data are obtained from 10 independent simulations of physical strings with $1024^3$ grids.
The vertical dotted line corresponds to $m_A \ell = 1$ and thus the loop collapse is suppressed in the right side of this line. Note that the typical initial size of the loop decoupled from the network is roughly the Hubble length, i.e. $\ell \sim \ell_H$. Hence the number of small loops cannot increase efficiently for $\ell_H > m_A^{-1}$ as one can clearly see in the local string case with $e=1$ (top panels). 
This is in contrast to the near-global string case, i.e. $e=0.01$ (bottom panels), where plenty of small loops are observed.
In that case, smaller loops are more abundant because the simulation box has more Hubble patches and thus more initial Hubble-size loops are produced at earlier times. Note that total number of strings decreases with time in any case because the number of independent Hubble patches in a simulation box decreases due to the cosmic expansion.

Let us consider the dark photon emission from the network of cosmic strings.
In the Abelian-Higgs model, the Nambu-Goldstone mode (i.e. the phase component) of the complex scalar field and the longitudinal mode of the gauge field after the symmetry breaking are related through the Gauss constraint (\ref{eq:GaussLaw}).\footnote{Eq.~(\ref{eq:GaussLaw}) has been derived in the temporal gauge $A_0=0$. For general gauge, it is sufficient to replace the time derivative of $\Phi$ with the covariant derivative.}
Therefore, in order to evaluate the energy density of the longitudinal mode, we should also sum up the Goldstone boson term. More explicitly, the energy density of the total system given by
\begin{align}
	\rho = \frac{1}{2a^4}(E_i^2 + B_i^2) +\frac{1}{a^2}\left( |D_0\Phi|^2 + |D_i \Phi|^2 \right)+ V.
\end{align}
To pick up the energy density of the ``longitudinal'' part $\rho^{({\rm L})}_A $, we must include both the $E_i^2$ term and the angular component of the $\Phi$ kinetic energy term. As a result, we obtain
\begin{align}
\rho^{({\rm L})}_A = \frac{|\Phi|^2}{v^2}\left[\frac{2}{a^2}\bigg(\frac{{\rm Im}(\Phi^* \Phi')}{|\Phi|}\bigg)^2+\frac{1}{a^4}\left(E^{({\rm L})}_i\right)^2 \right].
\label{rho_AL}
\end{align}
In order to screen the cosmic string and extract only the contribution of the free dark photon, the factor $|\Phi|/v$ is multiplied as in the axion string case \cite{Gorghetto:2018myk,Buschmann:2021sdq}. Namely, $|\Phi|/v \simeq 0$ near the cosmic sting and $|\Phi|/v=1$ far from the string core.
Note that the gradient term is highly contaminated by the cosmic string especially in the near-global string case (i.e. $e \ll 1$), so the inclusion the gradient term may cause the overestimate of the free dark photon abundance. Instead, factor 2 is multiplied in order to take into account the gradient term or mass term both of which would be comparable to the kinetic term.
By using the Gauss constraint (\ref{eq:GaussLaw}), the energy density (\ref{rho_AL}) is also rewritten as $\rho^{({\rm L})}_A \simeq \left[ \left(\partial_iE^{({\rm L})}_i\right)^2/(a^2 m_A^2)+  \left(E^{({\rm L})}_i\right)^2 \right]/a^4 $ for $|\Phi|\simeq v$. The first term, which was originally the Goldstone kinetic term, is dominant for $p\gg m_A$ with a typical physical momentum $p$. It is consistent with the intuition from the Goldstone equivalence theorem.

\begin{figure}
\begin{center}
   \includegraphics[width=8cm]{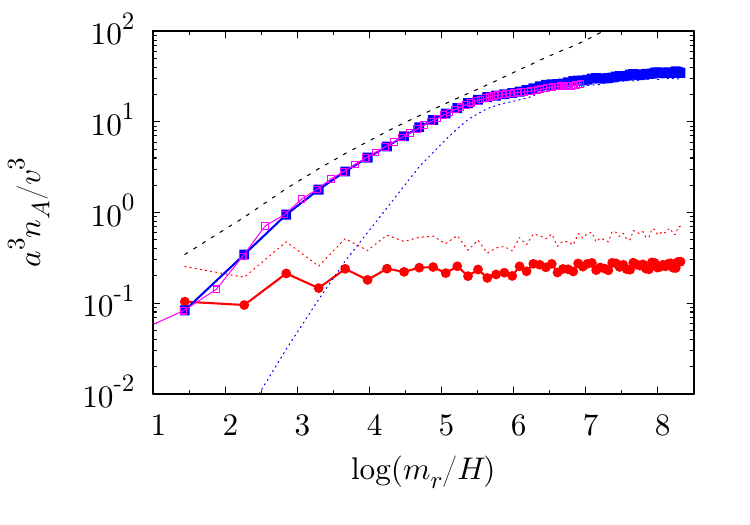}
    \includegraphics[width=8cm]{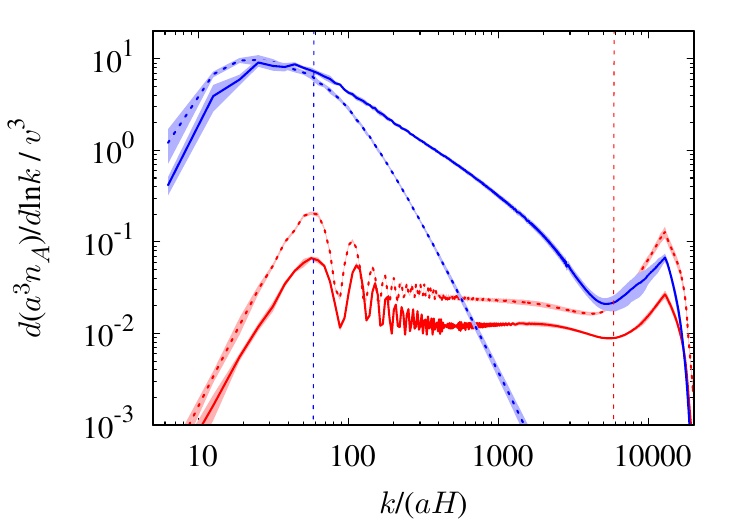}
    \includegraphics[width=8cm]{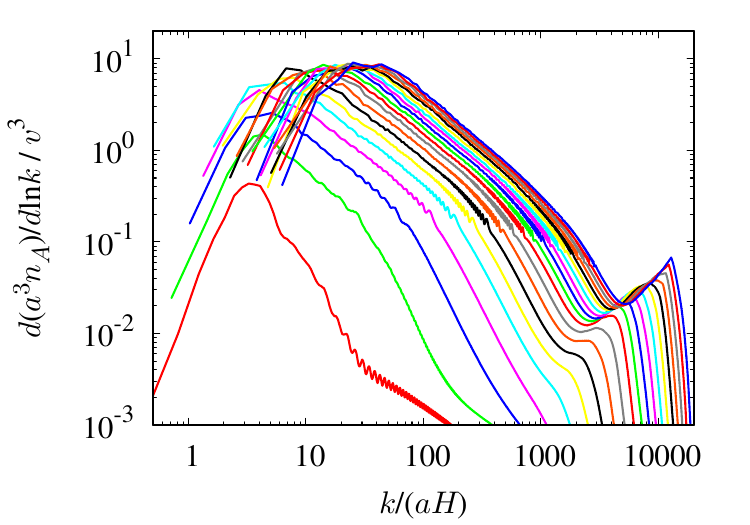}
  \end{center}
  \caption{Time evolution of the dark photon number density in a comoving box  (top-left), its spectrum at $\log(m_r/H)=8.3$ ($m_r\tau=64$) (top-right) with $\lambda = 2$ ($m_r=v$), $e=1$ (red), 0.01 (blue) and the evolution of the spectrum of the longitudinal mode (bottom) where time evolves from left to right ($m_r\tau=1$\,-\,64). We ran 5 simulations and the shaded region represents 1-$\sigma$ error. (The error bars are negligibly small in the left panel.) Thick solid lines and thin dotted lines correspond respectively to the longitudinal and transverse mode. In the top-left figure, The black-dashed line represents the analytic estimation given by Eq.~(\ref{eq:nAL}) with $\bar{E}_A/H=10$, $\mu=\pi v^2$ with numerically obtained values of $\xi$. Magenta line represents the result with twice larger simulation box size. In the right panel, horizontal axis is normalized by the horizon scale and each vertical dashed line represents $k/a= m_A$.}
  \label{fig:evolve_n}
\end{figure}

To estimate the final relic abundance of the dark photon, it is convenient to calculate the number density, $n_A$, rather than the energy density, since $n_A$ is conserved in a comoving volume. It can be calculated by
\begin{align}
n_A = \int dk \frac{dn_A}{dk} = \int dk \frac{1}{E_A(k)} \frac{d\rho_A}{dk}
\end{align}
with $E_A(k) = \sqrt{(k/a)^2+m_A^2}$.
The energy spectrum $d\rho_A/dk$ can be calculated by integrating the Fourier transformation of ${\rm Im}(\Phi^*\Phi')$ and $|\Phi|E_i$ with respect to the $k$-space solid angle.
Dark photons emitted from the collapse of loops have typically Hubble-scale momenta, $k/a \sim H$. On the other hand, the emission is kinematically suppressed for $H < m_A$. Thus, a typical energy of dark photons is $E_A(k) \sim H$ at the emission.
Since the energy density of emitted dark photons is proportional to that of the sting network and if the network follows the scaling law, one obtains the number density of newly produced dark photons at some cosmic time $t$ as $\Delta n_A \sim \rho_{\rm str}/H \propto 1/t$ and hence $a^3 \Delta n_A \propto t^{1/2}$. Therefore, such newly produced dark photons account for the dominant part of the total abundance, implying that the final dark photon abundance is determined by the number of dark photons produced at $H \sim m_A$.
When the dark photons are continuously produced from the collapse of string loops, the number density of longitudinal dark photons at each time slice can be roughly estimated as~\cite{Long:2019lwl}
\begin{align} \label{eq:nAL}
n_A^{({\rm L})}(t) \simeq \frac{8 \xi \mu H}{\bar{E}_A/H},
\end{align}
where $\bar{E}_A$ is the mean energy of produced dark photons.

Figure \ref{fig:evolve_n} shows the evolution (left) and spectrum (right) of the comoving number density of longitudinal (thick-solid lines) and transverse (thin-dotted lines) dark photons.
One can see that the dark photon production is not efficient for $e=1$. 
Persistent dark photon production is observed for $e=0.01$ until $\log(m_r/H) \sim 5$ but it stops afterward. 
This result can be well-fitted by Eq.~(\ref{eq:nAL}) with $\mu = \pi v^2$ and $\bar{E}_A/H=10$ (dashed-black line). 
The spectrum is mildly red-tilted ($\propto 1/k$), which explains the relative large value of $\bar{E}_A/H$.
Note that the transverse mode is also excited around $H \sim m_A$, and thus additional two polarization modes equally contribute to the total dark photon abundance. It is equivalent to multiplying $n_A^{({\rm L})}$ by a factor 3 to estimate the total $n_A$.
There are several ways to understand this. In the unitary gauge, the longitudinal boson coupling is enhanced by an amount $E_A/m_A$ due to the kinetic term normalization compared with the transverse one, whose coupling is just given by $e$. When $E_A\sim H \sim m_A$, they are comparable. In the Goldstone picture, the Goldstone mode $\theta$ couples like $\sim (\partial_\mu\theta/v)\,j^\mu$ while the transverse mode couples like $\sim eA_\mu^{\rm (T)}j^\mu$ where $j_\mu = 2\,{\rm Im}(\Phi^*\partial_\mu\Phi)$ is the U(1) current which only exists at the string core. Thus it is seen that the effective Goldstone coupling is $E_A /(ev) \sim E_A/m_A$ times that of the transverse mode.\footnote{
    As a result, the relative emission rate of GWs, longitudinal vector bosons and transverse vector bosons from the string loops is estimated as $P_{\rm GW} : P_{\rm L}: P_{\rm T} \sim G\mu : 1 : m_A^2/H^2$ for the string loop size $\ell \sim H^{-1} < m_A^{-1}$, where $G$ denotes the Newton constant. In the opposite case $\ell > m_A^{-1}$ we just assume that loops cannot emit a vector boson. See next section for estimation of the GW spectrum.
}
Therefore, the production of transverse vector boson is inefficient at early epoch, but eventually it is comparable to the longitudinal one.
This is in contrast to the statement in Ref.~\cite{Long:2019lwl}, where it has been claimed that the transverse mode production is negligible.

After the production stops, the number density of dark photons divided by the entropy density, $n_A/s$, is conserved if there is not any additional entropy production in the subsequent cosmological history. Then, one can obtain the final abundance of the dark photon
\begin{align}
\Omega_A h^2 = \frac{m_A (n_{A,0}/s_0) h^2}{\rho_{\rm cr,0}/s_0} \simeq 0.091 \bigg(\frac{\xi}{12} \bigg) \bigg(\frac{m_A}{10^{-13}\,{\rm eV}}\bigg)^{1/2} \bigg(\frac{v}{10^{14}\,{\rm GeV}}\bigg)^2,
\label{OmegaA}
\end{align}
where we have substituted $\bar{E}_A/H = 10$ obtained from the simulation (see Fig.~\ref{fig:evolve_n}). If there is no scaling violation, $\xi$ is constant and $\xi =3$\,-\,5 from our simulation with $e=0.01$. If one assumes the scaling violation, $\xi$ is given by the following fitting formula supported by the simulation of global strings in Refs.~\cite{Gorghetto:2018myk,Gorghetto:2020qws}, 
\begin{align}
\xi = 0.15\log\bigg(\frac{m_r}{m_A}\bigg) \simeq 12+0.15\log\left[ \bigg(\frac{m_r}{10^{14}\,{\rm GeV}}\bigg) \bigg(\frac{10^{-13}\,{\rm eV}}{m_A}\bigg) \right].
\label{xi}
\end{align}
Our simulation is performed with only a limited time range. To determine more precise values of the final dark photon DM abundance, we need more refined numerical simulations using more grid points. It is left for future work.

As a remark, we assumed the radiation-dominated Universe around $H=m_A$ in deriving Eq.~(\ref{OmegaA}). If the reheating temperature $T_{\rm R}$ is low enough, i.e. $T_{\rm R} \lesssim \sqrt{m_A M_{\rm Pl}}$ with $M_{\rm Pl}$ being the reduced Planck scale, the Universe might be inflaton oscillation dominated at $H=m_A$. In such a case, assuming that the inflaton oscillation behaves as non-relativistic matter, the final abundance should be multiplied by a factor $T_{\rm R}/\sqrt{m_A M_{\rm Pl}}$ to Eq.~(\ref{OmegaA}), up to some corrections to numerical factor, which indicates that the abundance becomes independent of $m_A$.

\section{Gravitational wave background from cosmic strings}  \label{sec:GW}

In this section we derive the spectrum of stochastic GW background from cosmic strings. In the dark photon DM scenario, a big difference from the case of local strings and pure global strings is that the dark photon is very light but still massive. The short loops that are created in the early Universe promptly decay through the dark photon emission and hence the GW emission efficiency is suppressed, while long loops that are created in relatively recent epoch cannot lose their energy through the dark photon emission. Thus the resulting GW spectrum will be significantly different from both the (pure) local string and global case.

\subsection{Energy loss of loops}

The energy density of the cosmic string network in the scaling regime is
\begin{align}
	\rho_{\rm str}(t) = \xi(t) \frac{\mu(t)}{t^2},  \label{rho_str}
\end{align}
where the $\xi$ parameter is given by (\ref{xi}). The string tension may be given by $\mu=2\pi v^2 \log(m_r/M)$ where $M\equiv {\rm max}[H, m_A]$. Within Hubble time, cosmic strings intersect with each other, forming loops which eventually decays emitting GWs or light vector bosons. 
The energy loss rate of a loop is given by
\begin{align}
	\frac{dE_{\ell}}{dt} = -\Gamma_{\rm GW} G\mu^2 - \Gamma_{\rm vec} v^2 \theta(1-m_A \ell),
\end{align}
where $\Gamma_{\rm GW} \sim 50$ and $\Gamma_{\rm vec}\sim 65$ are numerical constants~\cite{Vilenkin:1986ku,Vilenkin:2000jqa,Blanco-Pillado:2011egf,Blanco-Pillado:2013qja,Blanco-Pillado:2017rnf,Chang:2021afa}. The step function ensures that the loop cannot emit vector boson if the inverse loop size is smaller than the vector boson mass. 
For small loops $m_A \ell_i < 1$, the loop length at the time $t$ with an initial length $\ell_i$ is given by
\begin{align}
	\ell(t) = \ell_i - \Gamma_{\rm eff} (t-t_i),   \label{ellt}
\end{align}
where
\begin{align}
	 \Gamma_{\rm eff}  \equiv \Gamma_{\rm GW} G\mu + \frac{\Gamma_{\rm vec}}{2\pi \log(m_r/M)}.
\end{align}
For estimating the loop lifetime, the GW contribution is negligible unless the symmetry breaking $v$ is very close to the Planck scale. Thus the loop lifetime is
\begin{align}
	\tau(\ell_i) \simeq \frac{2\pi \log(m_r/M)}{\Gamma_{\rm vec}} \ell_i.
\end{align}
Up to a numerical factor of $\mathcal O(1)$, $\tau(\ell_i) \sim \ell_i$ in this case. Since $\ell_i < t_i$, the lifetime is shorter than the Hubble time. On the other hand, for large loops $m_A \ell_i > 1$, the GW emission is the only energy loss process\footnote{
	Still it can emit vector bosons from small scale irregularities on the loops. We do not consider it in this paper.
}
and we have
\begin{align}
	\tau(\ell_i) \simeq \frac{\ell_i}{\Gamma_{\rm GW}G\mu}.
\end{align}
Thus, for $\ell_i > \Gamma_{\rm GW} G\mu t_i$, the loops are long-lived, i.e., the lifetime is longer than the Hubble time.

\subsection{Loop density}

To maintain the scaling solution (\ref{rho_str}), the infinite strings cast their energy into string loops. An energy fraction $C_{\rm loop} \sim 10\,\%$ of the total string energy density is considered to go into the loops within one Hubble time. The loops lose their energy due to the emission of vector bosons or GWs. As far as $v\ll M_{\rm Pl}$ and the vector boson is much lighter than the inverse loop size $m_A \lesssim \ell^{-1}$, the emission of (longitudinal) vector boson is much more efficient than that into GWs. Even the Hubble-size loops decay within one Hubble time by the emission of vector bosons. Still, a small fraction of loop energy goes into GWs and also notice that loops larger than the inverse vector boson mass cannot emit vector boson and hence GW emission is the dominant source of the energy loss.
Below we estimate the stochastic GW background spectrum generated from the string loop decays.\footnote{
	GWs are also produced from infinite strings~\cite{Kawasaki:2010yi,Matsui:2016xnp,Matsui:2019obe}, which we will not pursue in this paper.
}

Let us denote the number density of string loops with a loop length $\ell$ per logarithmic bin of $\ell$ produced at the cosmic time $t$ as $\left[dn_\ell / d\ln \ell (t)\right]_{\rm prod}$. The total energy density of the string loops, averaged over the Hubble time, is
\begin{align}
	\rho_{\rm loop}(t) =\int\frac{t + \tau(\ell)}{\tau(\ell)}\left[\frac{d\rho_{\ell}(t)}{d\ln\ell}\right]_{\rm prod} d\ln\ell =  \int \mu \ell  \frac{t + \tau(\ell)}{\tau(\ell)} \left[\frac{dn_{\ell}(t)}{d\ln\ell}\right]_{\rm prod} d\ln\ell,
\end{align}
where the subscript ``prod'' indicates the energy/number density of loops that are produced at the time $t$ with the Hubble time interval and the factor $\frac{t + \tau(\ell)}{\tau(\ell)}$ is introduced taking account of the finite lifetime of the loop. If the loop lifetime $\tau(\ell)$ is shorter than the Hubble time $t$, loops exist only in a short fractional time interval $\tau(\ell) / t$ within one Hubble time.

The loops have some distributions on the initial loop length $\ell_i$. There are several studies on the initial loop size, but no definite conclusions have not been reached yet. Our simulation in Sec.~\ref{sec:sim} is also not enough to determine the loop distribution for very late Universe due to the limited dynamical range of the simulation. 
Instead, we consider typical possibilities for the initial distribution of $\alpha$. One is the monochromatic distribution, 
\begin{align}
	\left[\frac{d n_{\ell}(t)}{d\ln \ell}\right]_{\rm prod} =\frac{\tau(\ell)}{t + \tau(\ell)} \frac{C_{\rm loop} \xi}{\alpha_0 t^3} \,\ell\delta (\ell- \alpha_0 t),
	\label{mono}
\end{align}
where $\alpha_0=0.1$ will be taken. The overall normalization is determined such that $\rho_{\rm loop}(t) = C_{\rm loop} \rho_{\rm str}(t)$.
The other is the flat distribution in the log spacing of $\ell$ in the interval $ m_r^{-1} < \ell < \alpha_0 t$~\cite{Gorghetto:2018myk} (which we call ``log-flat distribution''),
\begin{align}
	\left[\frac{d n_{\ell}(t)}{d\ln \ell}\right]_{\rm prod} = 
	 \left[ \alpha_0 +\Gamma_{\rm eff} \log(\alpha_0 t m_r)\right]^{-1}  \frac{C_{\rm loop}\xi}{t^3}
	 \,\Theta(\ell; m_r^{-1}, \alpha_0 t) 
	\label{logflat}
\end{align}
where the function $\Theta(t; x,y)$ is defined as
\begin{align}
	\Theta(t; x,y) =\begin{cases}
	1 & {\rm for}~~x< t< y \\
	0 & {\rm otherwise}
	\end{cases}.
\end{align}
Again, the overall normalization is determined such that $\rho_{\rm loop}(t) = C_{\rm loop} \rho_{\rm str}(t)$.
If the vector boson emission is efficient we have $\tau(\ell) \sim \ell$ and if it is kinematically suppressed, the GW emission is dominant and we have $\tau(\ell) \gg \ell$. Thus the numerical factor in the parenthesis is always $\mathcal O(1)$.

After the formation of loops at $t=t_i$, their number density decrease as $a(t)^{-3}$ until they decay at $t=t_i + \tau(\ell)$. 
Thus the number density of loops of length $\ell$ measured at $t$ that were produced at $t=t_i$ is
\begin{align}
	\frac{d n_{\ell}(t;t_i)}{d\ln \ell} &= 
	\displaystyle \left[\frac{d n_{\ell}(t_i)}{d\ln \ell_i}\right]_{\rm prod} \left(\frac{a(t_i)}{a(t)}\right)^3 \theta\left(t-t_i\right),
	\label{dndell_ti}
\end{align}
where the initial loop size $\ell_i$ will evolve according to (\ref{ellt}).
The total loop number density of length $\ell$ at the time $t$ is then given by the integration over the production time $t_i$:
\begin{align}
	\frac{d n_{\ell}(t)}{d\ln \ell} = \int^t d\ln t_i \frac{d n_{\ell}(t;t_i)}{d\ln \ell_i}.
\end{align}

For the case of monochromatic distribution (\ref{mono}), the loop length $\ell_i$ and the production time $t_i$ has one-to-one correspondence as $\ell_i = \alpha_0 t_i$. We obtain
\begin{align}
	\frac{d n_{\ell}(t)}{d\ln \ell} \simeq
	\frac{\tau(\ell)}{\tau(\ell) + t_i} \frac{C_{\rm loop}\xi (t_i)}{\alpha_0 t_i^3}\,\left(\frac{a(t_i)}{a(t)}\right)^3 \theta\left(t-\frac{\ell_i}{\alpha_0}\right),
\end{align}
where
\begin{align}
	t_i = \frac{\ell + \Gamma_{\rm eff}t}{\alpha_0 + \Gamma_{\rm eff}}.
\end{align}
For the case of log-flat distribution (\ref{logflat}), by noting that the integral over $t_i$ is dominated by the lower edge $t_i \simeq \ell_i/\alpha_0$, we obtain
\begin{align}
	\frac{d n_{\ell}(t)}{d\ln \ell} \simeq
	\frac{1}{ \alpha_0 +\Gamma_{\rm eff} \log(\alpha_0 t m_r)} \frac{C_{\rm loop}\xi (t_i)}{t_i^3}\,\left(\frac{a(t_i)}{a(t)}\right)^3 \theta\left(t-\frac{\ell_i}{\alpha_0}\right).
\end{align}
By noting that the factor $\frac{\tau(\ell)}{\tau(\ell) + t_i}$ is at least $\mathcal O(0.1)$ in the monochromatic distribution with $\alpha_0\sim \mathcal O(0.1)$, we can see that the monochromatic and log-flat distributions do not give much differences for the purpose of orders-of-magnitude estimation. Below we use the monochromatic distribution for simplicity.


\subsection{Gravitational wave spectrum}

The GW power radiated from each loop with the size $\ell$ into the GW frequency $f$ is given by
\begin{align}
	\frac{d P_{\rm GW}(\ell)}{d \ln f} = G \mu^2 f \ell \, S(f\ell),  
\end{align}
where the spectral function $S(x)$ takes the form
\begin{align}
	S(x) = (q-1)2^{q-1}\Gamma_{\rm GW} \frac{\theta(x-2)}{x^q},  \label{Sx}
\end{align}
where $q>1$ is assumed and the overall normalization is determined so that $\int dx S(x) = \Gamma_{\rm GW} \sim 50$.
The slope $q$ depends on the mechanism of the GW emission. Such a power low tail in the high frequency part is produced by cusps or kinks on the loop. For cusps on the loop, we have $S(x)\propto x^{-4/3}$~\cite{Vachaspati:1984gt} and for kinks $S(x)\propto x^{-5/3}$~\cite{Garfinkle:1987yw,Damour:2001bk,Olmez:2010bi}. Since the cusp contribution is dominant, we take $q=4/3$ in the following.

The GW energy density with the frequency between $[f, f+df]$ produced by the loops with length between $[\ell, \ell+d\ell]$ per unit time at the time $t$ is given by
\begin{align}
	\frac{d\rho_{\rm GW}(t)}{dt} = \frac{d P_{\rm GW}(\ell)}{df}\times \frac{dn_\ell(t)}{d\ell} d\ell \times df(t).
\end{align}
The present GW frequency $f_0$ is given by $f_0 = f(t) a(t) / a_0$ and the energy density scales as $\rho_{\rm GW} \propto a^{-4}$. Thus the present GW spectrum is expressed as\footnote{
	To be precise, one may subtract GWs from ``rare burst'' events at relatively low redshift for estimating the stochastic GW background~\cite{Damour:2000wa,Damour:2001bk}.  
}
\begin{align}
	\frac{d\rho_{\rm GW}(t_0)}{d f_0} = \int dt \int d\ell \,G\mu^2(t)\,S\left(\frac{\ell f_0a_0}{a(t)}\right)\,\frac{dn_\ell(t)}{d\ln\ell} \left(\frac{a(t)}{a_0}\right)^3.
\end{align}
It is convenient to express it in terms of the density parameter 
\begin{align}
	\Omega_{\rm GW} (f_0) = \frac{1}{\rho_{\rm crit}} \frac{d\rho_{\rm GW}(t_0)}{d \ln f_0} .
\end{align}

\begin{figure}
\begin{center}
   \includegraphics[width=14cm]{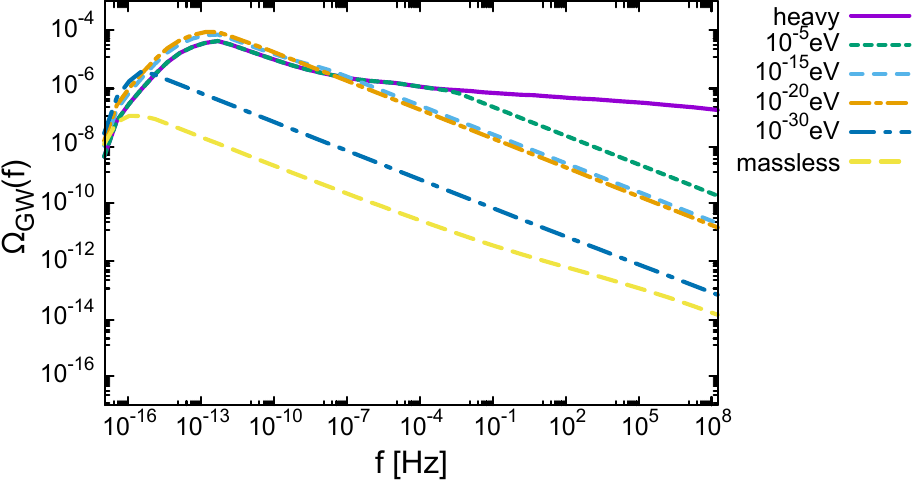}
   \vskip 1cm
   \includegraphics[width=14cm]{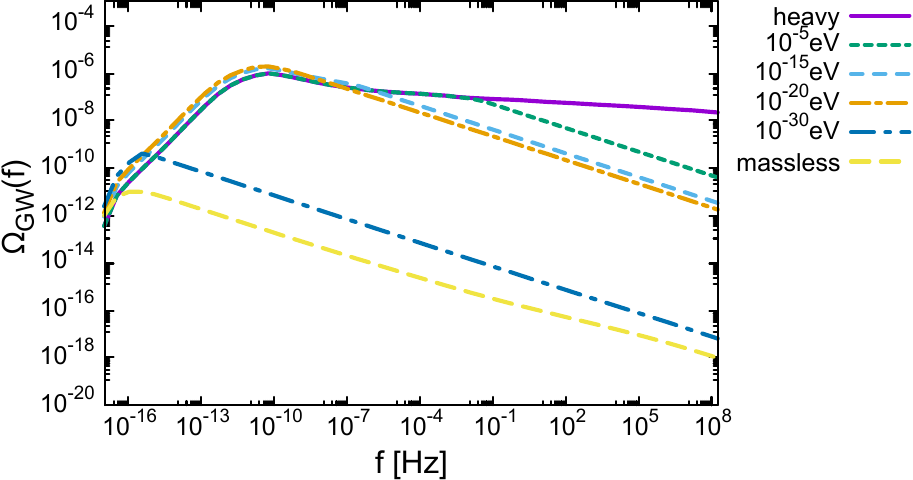}
  \end{center}
  \caption{The present day gravitational wave spectrum for $m_A=(\infty, 10^{-5}, 10^{-15}, 10^{-20}, 10^{-30}, 0)$\,eV, respectively. We have taken $v=10^{15}$\,GeV in the top panel and $v=10^{14}$\,GeV in the bottom panel. }
  \label{fig:gw}
\end{figure}

\begin{figure}
\begin{center}
   \includegraphics[width=14cm]{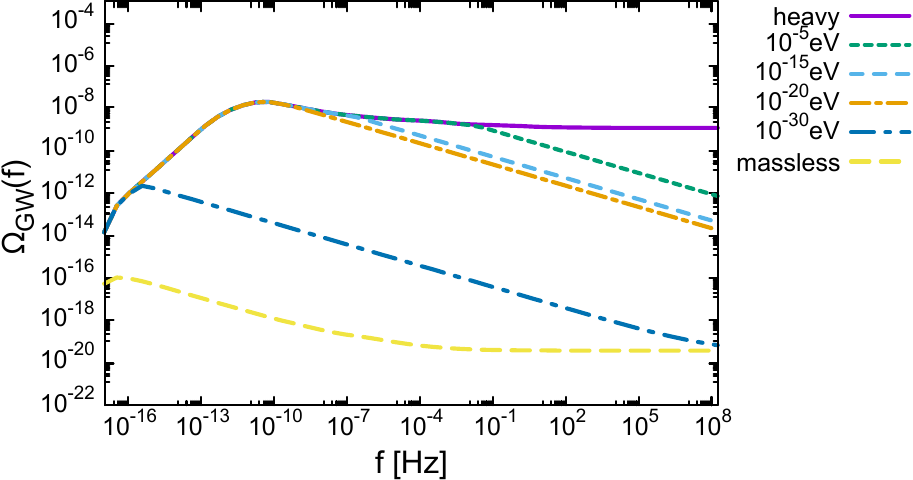}
   \vskip 1cm
   \includegraphics[width=14cm]{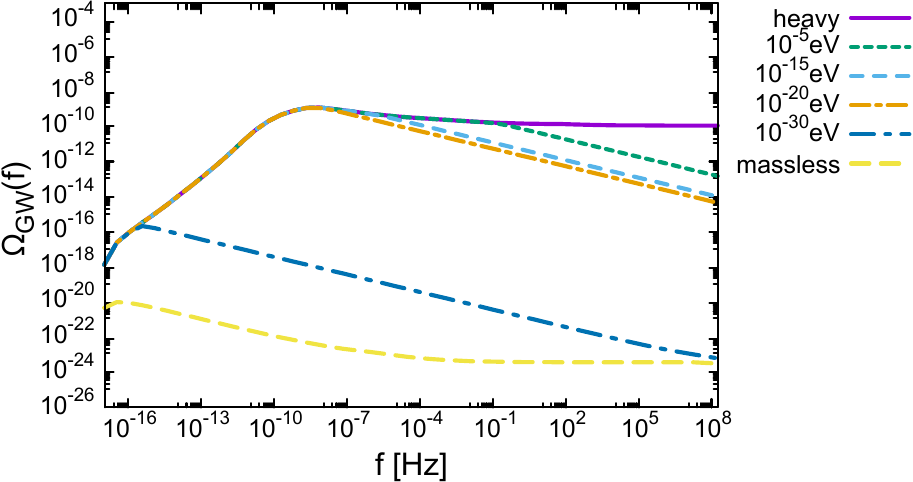}
  \end{center}
  \caption{The present day gravitational wave spectrum for $m_A=(\infty, 10^{-5}, 10^{-15}, 10^{-20}, 10^{-30}, 0)$\,eV, respectively. We have taken $v=10^{15}$\,GeV in the top panel and $v=10^{14}$\,GeV in the bottom panel. We have neglected all the log dependence by hand for illustrative purpose.}
  \label{fig:gw_nolog}
\end{figure}

Fig.~\ref{fig:gw} shows the present day gravitational wave spectrum for $m_A=(\infty,$ $10^{-5},$ $10^{-15},$ $10^{-20},$ $10^{-30},$ $ 0)$\,eV, respectively. We have taken $v=10^{15}$\,GeV in the top panel and $v=10^{14}$\,GeV in the bottom panel. 
For reference, $v=10^{15}$ $(10^{14})$\,GeV corresponds to $G\mu = 4\times 10^{-8}$ $(4\times 10^{-10})$ in terms of the frequently used parametrization $G\mu$ for the case of local string.
We have assumed monochromatic distribution for the loops. For illustrative purpose, we also show the spectrum with neglecting the log dependence by hand (i.e. $\xi=0.15$ and $\mu=2\pi v^2$) in Fig.~\ref{fig:gw_nolog}.
We have used the fitting formula given in Ref.~\cite{Saikawa:2018rcs} for the relativistic degrees of freedom $g_*$ and $g_{*s}$ for arbitrary redshift.

Let us explain a bit more detail. First let us see Fig.~\ref{fig:gw_nolog} where the log dependence is dropped by hand. 
\begin{itemize}
\item The case without vector boson emission (``heavy'' line in Fig.~\ref{fig:gw_nolog}) is just the same as the local string case. The flat part is roughly given by $\sim \sqrt{G\mu} \Omega_{\rm rad}$, since the fraction of loop energy density, which was created at $t=t_i$, to the total energy density is $\sim G\mu \times \left[a(\tau(\ell))/ a(t_i)\right] \sim \sqrt{G\mu}$ when the loop decays at $t=\tau(\ell)$ and all the energy of loops go to the GW emission.\footnote{
	Here we assumed that the loop was created and decays in the radiation dominated era. Note also that for very high frequency beyond the loop size that is created at the epoch of phase transition, the GW spectrum decreases as $f^{-1/3}$, which comes from the tail of the spectrum (\ref{Sx}).
} 
The low frequency limit of the GW spectrum is dominated by the loops with lifetime longer than the present age of the Universe, which emit GWs around $z\sim 1$. If $\alpha t_{\rm eq} < \Gamma_{\rm GW} G\mu t_0$, with $t_{\rm eq}$ being the cosmic time at the matter-radiation equality, all the loops that are created at the radiation dominated era completely decays until the present age. In this case, the low frequency spectrum is determined by the loops that are created at the matter domination era and it behaves as $\Omega_{\rm GW} \propto f$. In the opposite case, $\alpha t_{\rm eq} > \Gamma_{\rm GW} G\mu t_0$, loops that are created in the radiation dominated era can also exist in the present Universe and it gives a contribution like $\Omega_{\rm GW} \propto f^{3/2}$. For even lower frequency, loops that are created in the matter-dominated era contributes, which behaves as $\Omega_{\rm GW} \propto f$.
The peak of the spectrum corresponds to GWs from loops that are decaying in the present Universe. Such loops produce $f^{-1/3}$ tail towards higher frequency part until it crosses the flat part mentioned above. The peak frequency is roughly given by $f\sim (\Gamma_{\rm GW} G\mu t_0)^{-1}$ and the peak spectrum is given by $\Omega_{\rm GW}\sim G\mu$ for $\alpha t_{\rm eq} < \Gamma_{\rm GW} G\mu t_0$ and $\Omega_{\rm GW}\sim \sqrt{G\mu \alpha/\Gamma_{\rm GW}} \Omega_{\rm rad}^{3/4}$ for $\alpha t_{\rm eq} > \Gamma_{\rm GW} G\mu t_0$.

\item On the other hand, for the massless limit  (``massless'' line in Fig.~\ref{fig:gw_nolog}), the string loops are short-lived, i.e., they decay within one Hubble time. The overall normalization of the GW spectrum at the flat part is given by $\sim (G\mu)^2 \Omega_{\rm rad}$, since the most energy is consumed by the vector boson emission and only the small fraction $\sim G\mu$ of each loop goes to the GW emission. The lower frequency part is proportional to $f^{-1/3}$, which is a tail of the spectrum (\ref{Sx}) sourced by the loops of the size $\sim \alpha_0 t_0$, which are produced around $z\sim 1$.

\item By assuming finite vector boson mass $m_A$, there appear a break in the GW spectrum. For larger loops that are produced at later epochs the vector boson emission is prohibited and hence the GW spectrum in the low-frequency limit is similar to those in the massive limit. On the other hand, 
for smaller loops the GW emission is strongly suppressed because of the vector boson emission.
Thus the high frequency part is dominated by the $f^{-1/3}$ tail of the GW emission from large loops as described by the function (\ref{Sx}). Practically, small loops that are created in earlier epoch of the Universe decay into vector bosons and do not contribute to the present GW spectrum. It is hidden by the $f^{-1/3}$ tail coming from large long-lived loops that are decaying around $z\sim 1$.

\item The overall picture is similar for the case including the log dependence (Fig.~\ref{fig:gw}). Due to the log enhancement, the overall spectrum tilt is seen and the overall normalization also becomes significantly larger than the case without log dependence, since the string tension $\mu$ and $\xi$ parameter gets an log enhancement factor that directly increases the GW amplitude. The ``massless'' line of Fig.~\ref{fig:gw} is consistent with that given in Ref.~\cite{Chang:2021afa}, where the case of global strings with massless axion has been analyzed. 
\end{itemize}

\begin{figure}
\begin{center}
   \includegraphics[width=14cm]{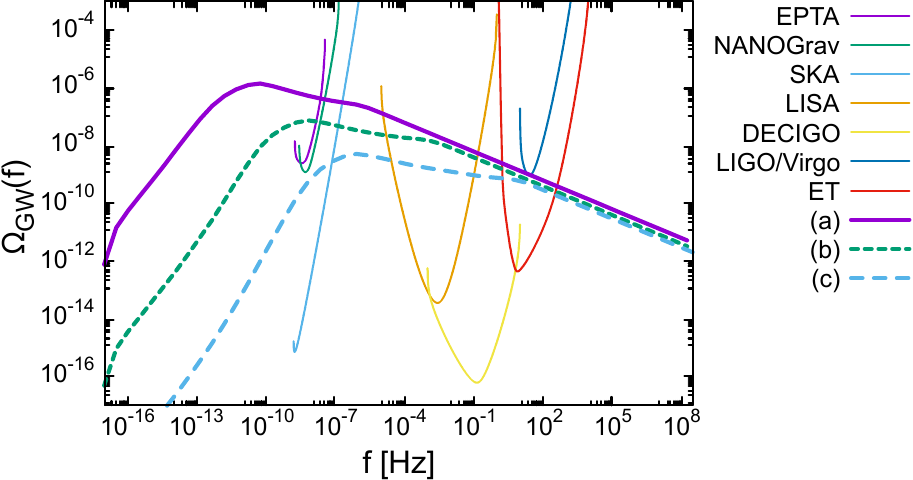}
   \vskip 1cm
   \includegraphics[width=14cm]{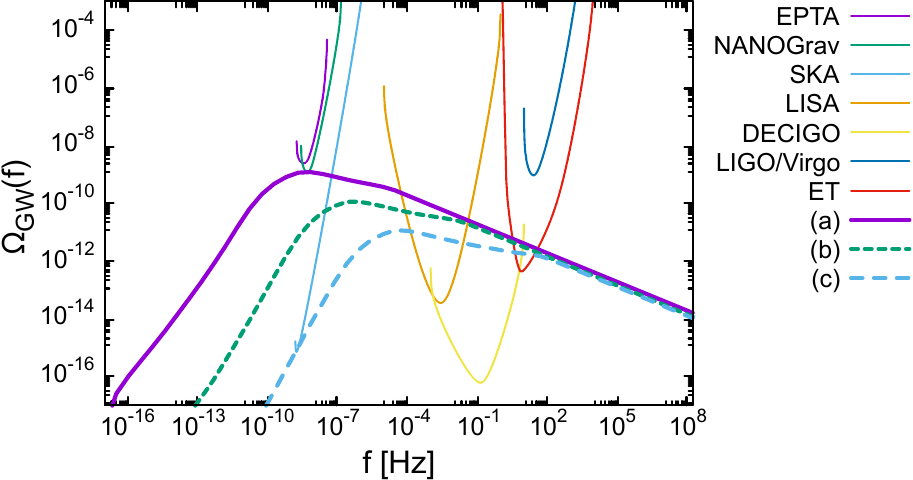}
  \end{center}
  \caption{(Top) GW spectrum from cosmic strings in the dark photon DM scenario, compared with current and future sensitivities of various experiments. Three theoretical lines correspond to benchmark points in which correct dark photon DM abundance is obtained: (a) $(v, m_A) = (10^{14}\,{\rm GeV}, 10^{-13}\,{\rm eV})$, (b) $(10^{13}\,{\rm GeV}, 10^{-9}\,{\rm eV})$, (c) $(10^{12}\,{\rm GeV}, 10^{-4}\,{\rm eV})$. (Bottom) The same as the top panel, but dropping the log dependence of $\mu$ and $\xi$ by hand.}
  \label{fig:sens}
\end{figure}

In Fig.~\ref{fig:sens} the calculated GW spectrum is compared with current and future sensitivities of various experiments: ground-based GW detectors such as LIGO~\cite{LIGOScientific:2017ikf,LIGOScientific:2014pky}, VIRGO~\cite{VIRGO:2014yos}, Einstein Telescope (ET)~\cite{Punturo:2010zz}, space laser interferometers such as LISA~\cite{Bartolo:2016ami}, DECIGO/BBO~\cite{Yagi:2011wg}, and pulsar timing arrays such as European Pulsar Timing Array (EPTA)~\cite{vanHaasteren:2011ni}, NANOGrav~\cite{NANOGrav:2020bcs}, Square Kilometer Array (SKA)~\cite{Janssen:2014dka}.
The sensitivity lines are taken from Refs.~\cite{Schmitz:2020syl,Schmitz:2020}. Note that lines of SKA, LISA, DECIGO/BBO and ET represent future sensitivities and others correspond to already existing bounds. 
We show three lines corresponding to benchmark points in which correct dark photon DM abundance is obtained within an order of magnitude (see Eq.~(\ref{OmegaA})): (a) $(v, m_A) = (10^{14}\,{\rm GeV}, 10^{-13}\,{\rm eV})$, (b) $(10^{13}\,{\rm GeV}, 10^{-9}\,{\rm eV})$, (c) $(10^{12}\,{\rm GeV}, 10^{-4}\,{\rm eV})$. 
The top panel assumes the log dependence of $\mu$ and $\xi$, while in the bottom panel we neglected the log dependence for comparison.

An interesting feature seen in Fig.~\ref{fig:sens} is that the three lines (a)-(c) converge in high-frequency limit. It is understood as follows. As explained above, the overall normalization of the GW spectrum is proportional to $\sqrt{G\mu}$ for long-lived loops. The spectral break appears at the frequency $f_{\rm break}$ that corresponds to the inverse of the loop size $\ell_i = \alpha_0 t_i \sim m_A^{-1}$, which decays at $\tau(\ell_i)\sim \ell_i/(\Gamma_{\rm GW}G\mu)$. The GW spectrum behaves as $\Omega_{\rm GW}\propto f^{-1/3}$ for $f> f_{\rm break}$. The break frequency is estimated as $f_{\rm break}\sim \ell_i^{-1}\left[a(\tau(\ell_i))/a_0 \right] \propto \sqrt{m_A/\mu}$ assuming that the corresponding loops are produced in the radiation-dominated era. On the other hand, the dark photon DM abundance is proportional to the combination $\mu\sqrt{m_A}$ (see Eq.~(\ref{OmegaA})). By requiring the consistent dark photon DM abundance, we obtain $f_{\rm break} \propto \mu^{-3/2}$. Therefore, we have $\Omega_{\rm GW}(f_{\rm break}) \propto f_{\rm break}^{-1/3}$. This power law dependence coincides with the $f^{-1/3}$ tail (\ref{Sx}). This is the reason why three lines (a)-(c) converge. It means that the GW amplitude at high frequency (e.g, in the DECIGO/BBO frequency range) is nearly independent of the dark photon mass if its abundance is consistent with the observed DM abundance. The parameter degeneracy is solved by combining with the low frequency observation, e.g. by SKA. 

It is seen that $v \gg 10^{12}$\,GeV or $m_A \ll 10^{-5}$\,eV is already excluded by the pulsar timing constraints if the log dependence of $\mu$ and $\xi$ are taken into account. 
Also it should be noticed that the on-going ground-based experiments, LIGO/Virgo, may be able to detect GWs independently of the dark photon mass.  
However, one should note that there may be several orders-of-magnitude uncertainties on the GW spectrum, especially from the log dependence of the string tension and $\xi$ parameter and also the loop abundance and such uncertainties significantly affect the resulting constraint on the symmetry breaking scale $v$ and the dark photon mass $m_A$. For example, if the GW spectrum is reduced by one order of magnitude due to these uncertainties, $m_A \sim 10^{-10}\,{\rm eV}$ may be allowed, as it is clearly seen in the bottom panel of Fig.~\ref{fig:sens}. Thus we need more accurate theoretical calculations to derive precise constraint.
In any case, future space laser-interferometers such as LISA and DECIGO/BBO, ground-based detectors such as ET and also the radio wave observations of pulsars by SKA have good chances to detect GWs from cosmic strings in the dark photon DM scenario.

\section{Conclusions and discussion} \label{sec:con}

Dark photon is one of the DM candidates and it is naturally produced by the cosmic string dynamics in association with the dark U(1) symmetry breaking. For a consistent DM abundance, dark photon mass may be very light and it implies that the dark gauge coupling constant should be very small. We first presented a field-theoretic simulation of cosmic strings with such a small gauge coupling. We confirmed that, within a precision of lattice simulation, the string dynamics is consistent with the naive expectation that the vector boson emission (or the Goldstone boson emission) is efficient for small loops $\ell \lesssim m_A^{-1}$ and the emission is suppressed otherwise.
Based on this observation, we calculated the stochastic GW background produced by the cosmic strings.
We found that the dark photon DM scenario from cosmic strings predict sizable amount of stochastic GW background, which may already be excluded by pulsar timing observations for $v \gg 10^{12}$\,GeV or $m_A \ll 10^{-5}$\,eV. The constraint on the string tension is stronger than the conventional local string case due to the logarithmic enhancement of the string tension and $\xi$ parameter. However, we need more precise study on this issue considering the large scale separation between $v$ and $m_A$ or the Hubble parameter, which is far beyond the reach of numerical simulation, since $\mathcal O(10)$ change in the GW spectrum would lead to orders-of-magnitude change to the constraint on $v$ or $m_A$.

Some notes are in order.
In calculating the GW spectrum, we made a simple assumption that loops longer than the inverse vector boson mass $m_A^{-1}$ cannot emit vector bosons. It may be a too simplified assumption since small scale irregularities may be present on long loops that allow to emit vector boson and long loops may lose their energy through the vector boson emission. Still we expect that the overall picture does not change drastically, since the energy of the order of $\mu \ell$ remains for the loop length $\ell$ even if such irregularities are smoothed out by the vector boson emission, and eventually it should be compensated by the GW emission. 
Another general remark is that the $f^{-1/3}$ tail of the GW spectrum represented by (\ref{Sx}) is quite important in broad frequency range as seen in figures. This power law behavior assumes that the cusp on the loop dominates the GW emission. While it is natural to consider that cusps take important roles since cusps are necessarily formed periodically on the string loop according to the Nambu-Goto action~\cite{Vilenkin:2000jqa}, still it requires more investigation whether it remains true in the case of our interest.
These are left for a future work.

\section*{Acknowledgment}

This work was supported by JSPS KAKENHI Grant Nos. 17H06359 (K.N.), 18K03609 (K.N.), 19H01894 (N.K.), 20H01894 (N.K.), 20H05851 (N.K.), 21H01078 (N.K.), 21KK0050 (N.K.).
This work was supported by World Premier International Research Center Initiative (WPI), MEXT, Japan.
Part of the results in this paper were obtained using supercomputing resources at Cyberscience Center, Tohoku University.

\bibliographystyle{utphys}
\bibliography{ref}

\providecommand{\href}[2]{#2}\begingroup\raggedright\begin{thebibliography}{100}

\bibitem{Fabbrichesi:2020wbt}
M.~Fabbrichesi, E.~Gabrielli, and G.~Lanfranchi, ``{The Dark Photon},''
  \href{http://arxiv.org/abs/2005.01515}{{\ttfamily arXiv:2005.01515
  [hep-ph]}}.

\bibitem{Caputo:2021eaa}
A.~Caputo, A.~J. Millar, C.~A.~J. O'Hare, and E.~Vitagliano, ``{Dark photon
  limits: A handbook},''
  \href{http://dx.doi.org/10.1103/PhysRevD.104.095029}{{\em Phys. Rev. D}
  {\bfseries 104} no.~9, (2021) 095029},
  \href{http://arxiv.org/abs/2105.04565}{{\ttfamily arXiv:2105.04565
  [hep-ph]}}.

\bibitem{Goodsell:2009xc}
M.~Goodsell, J.~Jaeckel, J.~Redondo, and A.~Ringwald, ``{Naturally Light Hidden
  Photons in LARGE Volume String Compactifications},''
  \href{http://dx.doi.org/10.1088/1126-6708/2009/11/027}{{\em JHEP} {\bfseries
  11} (2009) 027}, \href{http://arxiv.org/abs/0909.0515}{{\ttfamily
  arXiv:0909.0515 [hep-ph]}}.

\bibitem{Cicoli:2011yh}
M.~Cicoli, M.~Goodsell, J.~Jaeckel, and A.~Ringwald, ``{Testing String Vacua in
  the Lab: From a Hidden CMB to Dark Forces in Flux Compactifications},''
  \href{http://dx.doi.org/10.1007/JHEP07(2011)114}{{\em JHEP} {\bfseries 07}
  (2011) 114}, \href{http://arxiv.org/abs/1103.3705}{{\ttfamily arXiv:1103.3705
  [hep-th]}}.

\bibitem{Graham:2015rva}
P.~W. Graham, J.~Mardon, and S.~Rajendran, ``{Vector Dark Matter from
  Inflationary Fluctuations},''
  \href{http://dx.doi.org/10.1103/PhysRevD.93.103520}{{\em Phys. Rev. D}
  {\bfseries 93} no.~10, (2016) 103520},
  \href{http://arxiv.org/abs/1504.02102}{{\ttfamily arXiv:1504.02102
  [hep-ph]}}.

\bibitem{Ema:2019yrd}
Y.~Ema, K.~Nakayama, and Y.~Tang, ``{Production of purely gravitational dark
  matter: the case of fermion and vector boson},''
  \href{http://dx.doi.org/10.1007/JHEP07(2019)060}{{\em JHEP} {\bfseries 07}
  (2019) 060}, \href{http://arxiv.org/abs/1903.10973}{{\ttfamily
  arXiv:1903.10973 [hep-ph]}}.

\bibitem{Ahmed:2020fhc}
A.~Ahmed, B.~Grzadkowski, and A.~Socha, ``{Gravitational production of vector
  dark matter},'' \href{http://dx.doi.org/10.1007/JHEP08(2020)059}{{\em JHEP}
  {\bfseries 08} (2020) 059}, \href{http://arxiv.org/abs/2005.01766}{{\ttfamily
  arXiv:2005.01766 [hep-ph]}}.

\bibitem{Kolb:2020fwh}
E.~W. Kolb and A.~J. Long, ``{Completely dark photons from gravitational
  particle production during the inflationary era},''
  \href{http://dx.doi.org/10.1007/JHEP03(2021)283}{{\em JHEP} {\bfseries 03}
  (2021) 283}, \href{http://arxiv.org/abs/2009.03828}{{\ttfamily
  arXiv:2009.03828 [astro-ph.CO]}}.

\bibitem{Sato:2022jya}
T.~Sato, F.~Takahashi, and M.~Yamada, ``{Gravitational production of dark
  photon dark matter with mass generated by the Higgs mechanism},''
  \href{http://dx.doi.org/10.1088/1475-7516/2022/08/022}{{\em JCAP} {\bfseries
  08} no.~08, (2022) 022}, \href{http://arxiv.org/abs/2204.11896}{{\ttfamily
  arXiv:2204.11896 [hep-ph]}}.

\bibitem{Redi:2022zkt}
M.~Redi and A.~Tesi, ``{Dark photon Dark Matter without Stueckelberg mass},''
  \href{http://dx.doi.org/10.1007/JHEP10(2022)167}{{\em JHEP} {\bfseries 10}
  (2022) 167}, \href{http://arxiv.org/abs/2204.14274}{{\ttfamily
  arXiv:2204.14274 [hep-ph]}}.

\bibitem{Tang:2017hvq}
Y.~Tang and Y.-L. Wu, ``{On Thermal Gravitational Contribution to Particle
  Production and Dark Matter},''
  \href{http://dx.doi.org/10.1016/j.physletb.2017.10.034}{{\em Phys. Lett. B}
  {\bfseries 774} (2017) 676--681},
  \href{http://arxiv.org/abs/1708.05138}{{\ttfamily arXiv:1708.05138
  [hep-ph]}}.

\bibitem{Garny:2017kha}
M.~Garny, A.~Palessandro, M.~Sandora, and M.~S. Sloth, ``{Theory and
  Phenomenology of Planckian Interacting Massive Particles as Dark Matter},''
  \href{http://dx.doi.org/10.1088/1475-7516/2018/02/027}{{\em JCAP} {\bfseries
  02} (2018) 027}, \href{http://arxiv.org/abs/1709.09688}{{\ttfamily
  arXiv:1709.09688 [hep-ph]}}.

\bibitem{Agrawal:2018vin}
P.~Agrawal, N.~Kitajima, M.~Reece, T.~Sekiguchi, and F.~Takahashi, ``{Relic
  Abundance of Dark Photon Dark Matter},''
  \href{http://dx.doi.org/10.1016/j.physletb.2019.135136}{{\em Phys. Lett. B}
  {\bfseries 801} (2020) 135136},
  \href{http://arxiv.org/abs/1810.07188}{{\ttfamily arXiv:1810.07188
  [hep-ph]}}.

\bibitem{Co:2018lka}
R.~T. Co, A.~Pierce, Z.~Zhang, and Y.~Zhao, ``{Dark Photon Dark Matter Produced
  by Axion Oscillations},''
  \href{http://dx.doi.org/10.1103/PhysRevD.99.075002}{{\em Phys. Rev. D}
  {\bfseries 99} no.~7, (2019) 075002},
  \href{http://arxiv.org/abs/1810.07196}{{\ttfamily arXiv:1810.07196
  [hep-ph]}}.

\bibitem{Bastero-Gil:2018uel}
M.~Bastero-Gil, J.~Santiago, L.~Ubaldi, and R.~Vega-Morales, ``{Vector dark
  matter production at the end of inflation},''
  \href{http://dx.doi.org/10.1088/1475-7516/2019/04/015}{{\em JCAP} {\bfseries
  04} (2019) 015}, \href{http://arxiv.org/abs/1810.07208}{{\ttfamily
  arXiv:1810.07208 [hep-ph]}}.

\bibitem{Dror:2018pdh}
J.~A. Dror, K.~Harigaya, and V.~Narayan, ``{Parametric Resonance Production of
  Ultralight Vector Dark Matter},''
  \href{http://dx.doi.org/10.1103/PhysRevD.99.035036}{{\em Phys. Rev. D}
  {\bfseries 99} no.~3, (2019) 035036},
  \href{http://arxiv.org/abs/1810.07195}{{\ttfamily arXiv:1810.07195
  [hep-ph]}}.

\bibitem{Nakayama:2021avl}
K.~Nakayama and W.~Yin, ``{Hidden photon and axion dark matter from symmetry
  breaking},'' \href{http://dx.doi.org/10.1007/JHEP10(2021)026}{{\em JHEP}
  {\bfseries 10} (2021) 026}, \href{http://arxiv.org/abs/2105.14549}{{\ttfamily
  arXiv:2105.14549 [hep-ph]}}.

\bibitem{Long:2019lwl}
A.~J. Long and L.-T. Wang, ``{Dark Photon Dark Matter from a Network of Cosmic
  Strings},'' \href{http://dx.doi.org/10.1103/PhysRevD.99.063529}{{\em Phys.
  Rev. D} {\bfseries 99} no.~6, (2019) 063529},
  \href{http://arxiv.org/abs/1901.03312}{{\ttfamily arXiv:1901.03312
  [hep-ph]}}.

\bibitem{Nelson:2011sf}
A.~E. Nelson and J.~Scholtz, ``{Dark Light, Dark Matter and the Misalignment
  Mechanism},'' \href{http://dx.doi.org/10.1103/PhysRevD.84.103501}{{\em Phys.
  Rev. D} {\bfseries 84} (2011) 103501},
  \href{http://arxiv.org/abs/1105.2812}{{\ttfamily arXiv:1105.2812 [hep-ph]}}.

\bibitem{Arias:2012az}
P.~Arias, D.~Cadamuro, M.~Goodsell, J.~Jaeckel, J.~Redondo, and A.~Ringwald,
  ``{WISPy Cold Dark Matter},''
  \href{http://dx.doi.org/10.1088/1475-7516/2012/06/013}{{\em JCAP} {\bfseries
  06} (2012) 013}, \href{http://arxiv.org/abs/1201.5902}{{\ttfamily
  arXiv:1201.5902 [hep-ph]}}.

\bibitem{Nakayama:2019rhg}
K.~Nakayama, ``{Vector Coherent Oscillation Dark Matter},''
  \href{http://dx.doi.org/10.1088/1475-7516/2019/10/019}{{\em JCAP} {\bfseries
  10} (2019) 019}, \href{http://arxiv.org/abs/1907.06243}{{\ttfamily
  arXiv:1907.06243 [hep-ph]}}.

\bibitem{Nakayama:2020rka}
K.~Nakayama, ``{Constraint on Vector Coherent Oscillation Dark Matter with
  Kinetic Function},''
  \href{http://dx.doi.org/10.1088/1475-7516/2020/08/033}{{\em JCAP} {\bfseries
  08} (2020) 033}, \href{http://arxiv.org/abs/2004.10036}{{\ttfamily
  arXiv:2004.10036 [hep-ph]}}.

\bibitem{Nakai:2022dni}
Y.~Nakai, R.~Namba, and I.~Obata, ``{Peaky Production of Light Dark Photon Dark
  Matter},'' \href{http://arxiv.org/abs/2212.11516}{{\ttfamily arXiv:2212.11516
  [hep-ph]}}.

\bibitem{Vilenkin:2000jqa}
A.~Vilenkin and E.~P.~S. Shellard, {\em {Cosmic Strings and Other Topological
  Defects}}.
\newblock Cambridge University Press, 7, 2000.

\bibitem{Davis:1985pt}
R.~L. Davis, ``{Goldstone Bosons in String Models of Galaxy Formation},''
  \href{http://dx.doi.org/10.1103/PhysRevD.32.3172}{{\em Phys. Rev. D}
  {\bfseries 32} (1985) 3172}.

\bibitem{Davis:1986xc}
R.~L. Davis, ``{Cosmic Axions from Cosmic Strings},''
  \href{http://dx.doi.org/10.1016/0370-2693(86)90300-X}{{\em Phys. Lett. B}
  {\bfseries 180} (1986) 225--230}.

\bibitem{Vilenkin:1986ku}
A.~Vilenkin and T.~Vachaspati, ``{Radiation of Goldstone Bosons From Cosmic
  Strings},'' \href{http://dx.doi.org/10.1103/PhysRevD.35.1138}{{\em Phys. Rev.
  D} {\bfseries 35} (1987) 1138}.

\bibitem{Harari:1987ht}
D.~Harari and P.~Sikivie, ``{On the Evolution of Global Strings in the Early
  Universe},'' \href{http://dx.doi.org/10.1016/0370-2693(87)90032-3}{{\em Phys.
  Lett. B} {\bfseries 195} (1987) 361--365}.

\bibitem{Davis:1989nj}
R.~L. Davis and E.~P.~S. Shellard, ``{DO AXIONS NEED INFLATION?},''
  \href{http://dx.doi.org/10.1016/0550-3213(89)90187-9}{{\em Nucl. Phys. B}
  {\bfseries 324} (1989) 167--186}.

\bibitem{Dabholkar:1989ju}
A.~Dabholkar and J.~M. Quashnock, ``{Pinning Down the Axion},''
  \href{http://dx.doi.org/10.1016/0550-3213(90)90140-9}{{\em Nucl. Phys. B}
  {\bfseries 333} (1990) 815--832}.

\bibitem{Hagmann:1990mj}
C.~Hagmann and P.~Sikivie, ``{Computer simulations of the motion and decay of
  global strings},'' \href{http://dx.doi.org/10.1016/0550-3213(91)90243-Q}{{\em
  Nucl. Phys. B} {\bfseries 363} (1991) 247--280}.

\bibitem{Battye:1993jv}
R.~A. Battye and E.~P.~S. Shellard, ``{Global string radiation},''
  \href{http://dx.doi.org/10.1016/0550-3213(94)90573-8}{{\em Nucl. Phys. B}
  {\bfseries 423} (1994) 260--304},
  \href{http://arxiv.org/abs/astro-ph/9311017}{{\ttfamily
  arXiv:astro-ph/9311017}}.

\bibitem{Battye:1994au}
R.~A. Battye and E.~P.~S. Shellard, ``{Axion string constraints},''
  \href{http://dx.doi.org/10.1103/PhysRevLett.73.2954}{{\em Phys. Rev. Lett.}
  {\bfseries 73} (1994) 2954--2957},
  \href{http://arxiv.org/abs/astro-ph/9403018}{{\ttfamily
  arXiv:astro-ph/9403018}}. [Erratum: Phys.Rev.Lett. 76, 2203--2204 (1996)].

\bibitem{Yamaguchi:1998gx}
M.~Yamaguchi, M.~Kawasaki, and J.~Yokoyama, ``{Evolution of axionic strings and
  spectrum of axions radiated from them},''
  \href{http://dx.doi.org/10.1103/PhysRevLett.82.4578}{{\em Phys. Rev. Lett.}
  {\bfseries 82} (1999) 4578--4581},
  \href{http://arxiv.org/abs/hep-ph/9811311}{{\ttfamily arXiv:hep-ph/9811311}}.

\bibitem{Yamaguchi:1999yp}
M.~Yamaguchi, ``{Scaling property of the global string in the radiation
  dominated universe},''
  \href{http://dx.doi.org/10.1103/PhysRevD.60.103511}{{\em Phys. Rev. D}
  {\bfseries 60} (1999) 103511},
  \href{http://arxiv.org/abs/hep-ph/9907506}{{\ttfamily arXiv:hep-ph/9907506}}.

\bibitem{Yamaguchi:1999dy}
M.~Yamaguchi, J.~Yokoyama, and M.~Kawasaki, ``{Evolution of a global string
  network in a matter dominated universe},''
  \href{http://dx.doi.org/10.1103/PhysRevD.61.061301}{{\em Phys. Rev. D}
  {\bfseries 61} (2000) 061301},
  \href{http://arxiv.org/abs/hep-ph/9910352}{{\ttfamily arXiv:hep-ph/9910352}}.

\bibitem{Hagmann:2000ja}
C.~Hagmann, S.~Chang, and P.~Sikivie, ``{Axion radiation from strings},''
  \href{http://dx.doi.org/10.1103/PhysRevD.63.125018}{{\em Phys. Rev. D}
  {\bfseries 63} (2001) 125018},
  \href{http://arxiv.org/abs/hep-ph/0012361}{{\ttfamily arXiv:hep-ph/0012361}}.

\bibitem{Hiramatsu:2010yu}
T.~Hiramatsu, M.~Kawasaki, T.~Sekiguchi, M.~Yamaguchi, and J.~Yokoyama,
  ``{Improved estimation of radiated axions from cosmological axionic
  strings},'' \href{http://dx.doi.org/10.1103/PhysRevD.83.123531}{{\em Phys.
  Rev. D} {\bfseries 83} (2011) 123531},
  \href{http://arxiv.org/abs/1012.5502}{{\ttfamily arXiv:1012.5502 [hep-ph]}}.

\bibitem{Hiramatsu:2012gg}
T.~Hiramatsu, M.~Kawasaki, K.~Saikawa, and T.~Sekiguchi, ``{Production of dark
  matter axions from collapse of string-wall systems},''
  \href{http://dx.doi.org/10.1103/PhysRevD.85.105020}{{\em Phys. Rev. D}
  {\bfseries 85} (2012) 105020},
  \href{http://arxiv.org/abs/1202.5851}{{\ttfamily arXiv:1202.5851 [hep-ph]}}.
  [Erratum: Phys.Rev.D 86, 089902 (2012)].

\bibitem{Fleury:2015aca}
L.~Fleury and G.~D. Moore, ``{Axion dark matter: strings and their cores},''
  \href{http://dx.doi.org/10.1088/1475-7516/2016/01/004}{{\em JCAP} {\bfseries
  01} (2016) 004}, \href{http://arxiv.org/abs/1509.00026}{{\ttfamily
  arXiv:1509.00026 [hep-ph]}}.

\bibitem{Klaer:2017qhr}
V.~B. Klaer and G.~D. Moore, ``{How to simulate global cosmic strings with
  large string tension},''
  \href{http://dx.doi.org/10.1088/1475-7516/2017/10/043}{{\em JCAP} {\bfseries
  10} (2017) 043}, \href{http://arxiv.org/abs/1707.05566}{{\ttfamily
  arXiv:1707.05566 [hep-ph]}}.

\bibitem{Gorghetto:2018myk}
M.~Gorghetto, E.~Hardy, and G.~Villadoro, ``{Axions from Strings: the
  Attractive Solution},'' \href{http://dx.doi.org/10.1007/JHEP07(2018)151}{{\em
  JHEP} {\bfseries 07} (2018) 151},
  \href{http://arxiv.org/abs/1806.04677}{{\ttfamily arXiv:1806.04677
  [hep-ph]}}.

\bibitem{Kawasaki:2018bzv}
M.~Kawasaki, T.~Sekiguchi, M.~Yamaguchi, and J.~Yokoyama, ``{Long-term dynamics
  of cosmological axion strings},''
  \href{http://dx.doi.org/10.1093/ptep/pty098}{{\em PTEP} {\bfseries 2018}
  no.~9, (2018) 091E01}, \href{http://arxiv.org/abs/1806.05566}{{\ttfamily
  arXiv:1806.05566 [hep-ph]}}.

\bibitem{Buschmann:2019icd}
M.~Buschmann, J.~W. Foster, and B.~R. Safdi, ``{Early-Universe Simulations of
  the Cosmological Axion},''
  \href{http://dx.doi.org/10.1103/PhysRevLett.124.161103}{{\em Phys. Rev.
  Lett.} {\bfseries 124} no.~16, (2020) 161103},
  \href{http://arxiv.org/abs/1906.00967}{{\ttfamily arXiv:1906.00967
  [astro-ph.CO]}}.

\bibitem{Hindmarsh:2019csc}
M.~Hindmarsh, J.~Lizarraga, A.~Lopez-Eiguren, and J.~Urrestilla, ``{Scaling
  Density of Axion Strings},''
  \href{http://dx.doi.org/10.1103/PhysRevLett.124.021301}{{\em Phys. Rev.
  Lett.} {\bfseries 124} no.~2, (2020) 021301},
  \href{http://arxiv.org/abs/1908.03522}{{\ttfamily arXiv:1908.03522
  [astro-ph.CO]}}.

\bibitem{Klaer:2019fxc}
V.~B. Klaer and G.~D. Moore, ``{Global cosmic string networks as a function of
  tension},'' \href{http://dx.doi.org/10.1088/1475-7516/2020/06/021}{{\em JCAP}
  {\bfseries 06} (2020) 021}, \href{http://arxiv.org/abs/1912.08058}{{\ttfamily
  arXiv:1912.08058 [hep-ph]}}.

\bibitem{Gorghetto:2020qws}
M.~Gorghetto, E.~Hardy, and G.~Villadoro, ``{More axions from strings},''
  \href{http://dx.doi.org/10.21468/SciPostPhys.10.2.050}{{\em SciPost Phys.}
  {\bfseries 10} no.~2, (2021) 050},
  \href{http://arxiv.org/abs/2007.04990}{{\ttfamily arXiv:2007.04990
  [hep-ph]}}.

\bibitem{Hindmarsh:2021vih}
M.~Hindmarsh, J.~Lizarraga, A.~Lopez-Eiguren, and J.~Urrestilla, ``{Approach to
  scaling in axion string networks},''
  \href{http://dx.doi.org/10.1103/PhysRevD.103.103534}{{\em Phys. Rev. D}
  {\bfseries 103} no.~10, (2021) 103534},
  \href{http://arxiv.org/abs/2102.07723}{{\ttfamily arXiv:2102.07723
  [astro-ph.CO]}}.

\bibitem{Buschmann:2021sdq}
M.~Buschmann, J.~W. Foster, A.~Hook, A.~Peterson, D.~E. Willcox, W.~Zhang, and
  B.~R. Safdi, ``{Dark matter from axion strings with adaptive mesh
  refinement},'' \href{http://dx.doi.org/10.1038/s41467-022-28669-y}{{\em
  Nature Commun.} {\bfseries 13} no.~1, (2022) 1049},
  \href{http://arxiv.org/abs/2108.05368}{{\ttfamily arXiv:2108.05368
  [hep-ph]}}.

\bibitem{Blanco-Pillado:2022axf}
J.~J. Blanco-Pillado, D.~Jim\'enez-Aguilar, J.~M. Queiruga, and J.~Urrestilla,
  ``{Parametric Resonances in Axionic Cosmic Strings},''
  \href{http://arxiv.org/abs/2212.06194}{{\ttfamily arXiv:2212.06194
  [hep-th]}}.

\bibitem{Vachaspati:1984gt}
T.~Vachaspati and A.~Vilenkin, ``{Gravitational Radiation from Cosmic
  Strings},'' \href{http://dx.doi.org/10.1103/PhysRevD.31.3052}{{\em Phys. Rev.
  D} {\bfseries 31} (1985) 3052}.

\bibitem{Garfinkle:1987yw}
D.~Garfinkle and T.~Vachaspati, ``{Radiation From Kinky, Cuspless Cosmic
  Loops},'' \href{http://dx.doi.org/10.1103/PhysRevD.36.2229}{{\em Phys. Rev.
  D} {\bfseries 36} (1987) 2229}.

\bibitem{Caldwell:1991jj}
R.~R. Caldwell and B.~Allen, ``{Cosmological constraints on cosmic string
  gravitational radiation},''
  \href{http://dx.doi.org/10.1103/PhysRevD.45.3447}{{\em Phys. Rev. D}
  {\bfseries 45} (1992) 3447--3468}.

\bibitem{Caldwell:1996en}
R.~R. Caldwell, R.~A. Battye, and E.~P.~S. Shellard, ``{Relic gravitational
  waves from cosmic strings: Updated constraints and opportunities for
  detection},'' \href{http://dx.doi.org/10.1103/PhysRevD.54.7146}{{\em Phys.
  Rev. D} {\bfseries 54} (1996) 7146--7152},
  \href{http://arxiv.org/abs/astro-ph/9607130}{{\ttfamily
  arXiv:astro-ph/9607130}}.

\bibitem{Damour:2000wa}
T.~Damour and A.~Vilenkin, ``{Gravitational wave bursts from cosmic strings},''
  \href{http://dx.doi.org/10.1103/PhysRevLett.85.3761}{{\em Phys. Rev. Lett.}
  {\bfseries 85} (2000) 3761--3764},
  \href{http://arxiv.org/abs/gr-qc/0004075}{{\ttfamily arXiv:gr-qc/0004075}}.

\bibitem{Damour:2001bk}
T.~Damour and A.~Vilenkin, ``{Gravitational wave bursts from cusps and kinks on
  cosmic strings},'' \href{http://dx.doi.org/10.1103/PhysRevD.64.064008}{{\em
  Phys. Rev. D} {\bfseries 64} (2001) 064008},
  \href{http://arxiv.org/abs/gr-qc/0104026}{{\ttfamily arXiv:gr-qc/0104026}}.

\bibitem{Damour:2004kw}
T.~Damour and A.~Vilenkin, ``{Gravitational radiation from cosmic
  (super)strings: Bursts, stochastic background, and observational windows},''
  \href{http://dx.doi.org/10.1103/PhysRevD.71.063510}{{\em Phys. Rev. D}
  {\bfseries 71} (2005) 063510},
  \href{http://arxiv.org/abs/hep-th/0410222}{{\ttfamily arXiv:hep-th/0410222}}.

\bibitem{Siemens:2006yp}
X.~Siemens, V.~Mandic, and J.~Creighton, ``{Gravitational wave stochastic
  background from cosmic (super)strings},''
  \href{http://dx.doi.org/10.1103/PhysRevLett.98.111101}{{\em Phys. Rev. Lett.}
  {\bfseries 98} (2007) 111101},
  \href{http://arxiv.org/abs/astro-ph/0610920}{{\ttfamily
  arXiv:astro-ph/0610920}}.

\bibitem{DePies:2007bm}
M.~R. DePies and C.~J. Hogan, ``{Stochastic Gravitational Wave Background from
  Light Cosmic Strings},''
  \href{http://dx.doi.org/10.1103/PhysRevD.75.125006}{{\em Phys. Rev. D}
  {\bfseries 75} (2007) 125006},
  \href{http://arxiv.org/abs/astro-ph/0702335}{{\ttfamily
  arXiv:astro-ph/0702335}}.

\bibitem{Olmez:2010bi}
S.~Olmez, V.~Mandic, and X.~Siemens, ``{Gravitational-Wave Stochastic
  Background from Kinks and Cusps on Cosmic Strings},''
  \href{http://dx.doi.org/10.1103/PhysRevD.81.104028}{{\em Phys. Rev. D}
  {\bfseries 81} (2010) 104028},
  \href{http://arxiv.org/abs/1004.0890}{{\ttfamily arXiv:1004.0890
  [astro-ph.CO]}}.

\bibitem{Binetruy:2012ze}
P.~Binetruy, A.~Bohe, C.~Caprini, and J.-F. Dufaux, ``{Cosmological Backgrounds
  of Gravitational Waves and eLISA/NGO: Phase Transitions, Cosmic Strings and
  Other Sources},'' \href{http://dx.doi.org/10.1088/1475-7516/2012/06/027}{{\em
  JCAP} {\bfseries 06} (2012) 027},
  \href{http://arxiv.org/abs/1201.0983}{{\ttfamily arXiv:1201.0983 [gr-qc]}}.

\bibitem{Kuroyanagi:2012wm}
S.~Kuroyanagi, K.~Miyamoto, T.~Sekiguchi, K.~Takahashi, and J.~Silk,
  ``{Forecast constraints on cosmic string parameters from gravitational wave
  direct detection experiments},''
  \href{http://dx.doi.org/10.1103/PhysRevD.86.023503}{{\em Phys. Rev. D}
  {\bfseries 86} (2012) 023503},
  \href{http://arxiv.org/abs/1202.3032}{{\ttfamily arXiv:1202.3032
  [astro-ph.CO]}}.

\bibitem{Kuroyanagi:2012jf}
S.~Kuroyanagi, K.~Miyamoto, T.~Sekiguchi, K.~Takahashi, and J.~Silk,
  ``{Forecast constraints on cosmic strings from future CMB, pulsar timing and
  gravitational wave direct detection experiments},''
  \href{http://dx.doi.org/10.1103/PhysRevD.87.023522}{{\em Phys. Rev. D}
  {\bfseries 87} no.~2, (2013) 023522},
  \href{http://arxiv.org/abs/1210.2829}{{\ttfamily arXiv:1210.2829
  [astro-ph.CO]}}. [Erratum: Phys.Rev.D 87, 069903 (2013)].

\bibitem{Ringeval:2017eww}
C.~Ringeval and T.~Suyama, ``{Stochastic gravitational waves from cosmic string
  loops in scaling},''
  \href{http://dx.doi.org/10.1088/1475-7516/2017/12/027}{{\em JCAP} {\bfseries
  12} (2017) 027}, \href{http://arxiv.org/abs/1709.03845}{{\ttfamily
  arXiv:1709.03845 [astro-ph.CO]}}.

\bibitem{Cui:2017ufi}
Y.~Cui, M.~Lewicki, D.~E. Morrissey, and J.~D. Wells, ``{Cosmic Archaeology
  with Gravitational Waves from Cosmic Strings},''
  \href{http://dx.doi.org/10.1103/PhysRevD.97.123505}{{\em Phys. Rev. D}
  {\bfseries 97} no.~12, (2018) 123505},
  \href{http://arxiv.org/abs/1711.03104}{{\ttfamily arXiv:1711.03104
  [hep-ph]}}.

\bibitem{Gouttenoire:2019kij}
Y.~Gouttenoire, G.~Servant, and P.~Simakachorn, ``{Beyond the Standard Models
  with Cosmic Strings},''
  \href{http://dx.doi.org/10.1088/1475-7516/2020/07/032}{{\em JCAP} {\bfseries
  07} (2020) 032}, \href{http://arxiv.org/abs/1912.02569}{{\ttfamily
  arXiv:1912.02569 [hep-ph]}}.

\bibitem{Gorghetto:2021fsn}
M.~Gorghetto, E.~Hardy, and H.~Nicolaescu, ``{Observing invisible axions with
  gravitational waves},''
  \href{http://dx.doi.org/10.1088/1475-7516/2021/06/034}{{\em JCAP} {\bfseries
  06} (2021) 034}, \href{http://arxiv.org/abs/2101.11007}{{\ttfamily
  arXiv:2101.11007 [hep-ph]}}.

\bibitem{Chang:2021afa}
C.-F. Chang and Y.~Cui, ``{Gravitational waves from global cosmic strings and
  cosmic archaeology},'' \href{http://dx.doi.org/10.1007/JHEP03(2022)114}{{\em
  JHEP} {\bfseries 03} (2022) 114},
  \href{http://arxiv.org/abs/2106.09746}{{\ttfamily arXiv:2106.09746
  [hep-ph]}}.

\bibitem{Blanco-Pillado:1998tyu}
J.~J. Blanco-Pillado and K.~D. Olum, ``{Form of cosmic string cusps},''
  \href{http://dx.doi.org/10.1103/PhysRevD.59.063508}{{\em Phys. Rev. D}
  {\bfseries 59} (1999) 063508},
  \href{http://arxiv.org/abs/gr-qc/9810005}{{\ttfamily arXiv:gr-qc/9810005}}.
  [Erratum: Phys.Rev.D 103, 029902 (2021)].

\bibitem{Olum:1998ag}
K.~D. Olum and J.~J. Blanco-Pillado, ``{Field theory simulation of Abelian
  Higgs cosmic string cusps},''
  \href{http://dx.doi.org/10.1103/PhysRevD.60.023503}{{\em Phys. Rev. D}
  {\bfseries 60} (1999) 023503},
  \href{http://arxiv.org/abs/gr-qc/9812040}{{\ttfamily arXiv:gr-qc/9812040}}.

\bibitem{Hindmarsh:2017qff}
M.~Hindmarsh, J.~Lizarraga, J.~Urrestilla, D.~Daverio, and M.~Kunz, ``{Scaling
  from gauge and scalar radiation in Abelian Higgs string networks},''
  \href{http://dx.doi.org/10.1103/PhysRevD.96.023525}{{\em Phys. Rev. D}
  {\bfseries 96} no.~2, (2017) 023525},
  \href{http://arxiv.org/abs/1703.06696}{{\ttfamily arXiv:1703.06696
  [astro-ph.CO]}}.

\bibitem{Matsunami:2019fss}
D.~Matsunami, L.~Pogosian, A.~Saurabh, and T.~Vachaspati, ``{Decay of Cosmic
  String Loops Due to Particle Radiation},''
  \href{http://dx.doi.org/10.1103/PhysRevLett.122.201301}{{\em Phys. Rev.
  Lett.} {\bfseries 122} no.~20, (2019) 201301},
  \href{http://arxiv.org/abs/1903.05102}{{\ttfamily arXiv:1903.05102
  [hep-ph]}}.

\bibitem{Saurabh:2020pqe}
A.~Saurabh, T.~Vachaspati, and L.~Pogosian, ``{Decay of Cosmic Global String
  Loops},'' \href{http://dx.doi.org/10.1103/PhysRevD.101.083522}{{\em Phys.
  Rev. D} {\bfseries 101} no.~8, (2020) 083522},
  \href{http://arxiv.org/abs/2001.01030}{{\ttfamily arXiv:2001.01030
  [hep-ph]}}.

\bibitem{Jeannerot:1999yn}
R.~Jeannerot, X.~Zhang, and R.~H. Brandenberger, ``{Non-thermal production of
  neutralino cold dark matter from cosmic string decays},''
  \href{http://dx.doi.org/10.1088/1126-6708/1999/12/003}{{\em JHEP} {\bfseries
  12} (1999) 003}, \href{http://arxiv.org/abs/hep-ph/9901357}{{\ttfamily
  arXiv:hep-ph/9901357}}.

\bibitem{Cui:2007js}
Y.~Cui, S.~P. Martin, D.~E. Morrissey, and J.~D. Wells, ``{Cosmic Strings from
  Supersymmetric Flat Directions},''
  \href{http://dx.doi.org/10.1103/PhysRevD.77.043528}{{\em Phys. Rev. D}
  {\bfseries 77} (2008) 043528},
  \href{http://arxiv.org/abs/0709.0950}{{\ttfamily arXiv:0709.0950 [hep-ph]}}.

\bibitem{Cui:2008bd}
Y.~Cui and D.~E. Morrissey, ``{Non-Thermal Dark Matter from Cosmic Strings},''
  \href{http://dx.doi.org/10.1103/PhysRevD.79.083532}{{\em Phys. Rev. D}
  {\bfseries 79} (2009) 083532},
  \href{http://arxiv.org/abs/0805.1060}{{\ttfamily arXiv:0805.1060 [hep-ph]}}.

\bibitem{Kawasaki:2011dp}
M.~Kawasaki, K.~Miyamoto, and K.~Nakayama, ``{Cosmological Effects of Decaying
  Cosmic String Loops with TeV Scale Width},''
  \href{http://arxiv.org/abs/1105.4383}{{\ttfamily arXiv:1105.4383 [hep-ph]}}.

\bibitem{Miyamoto:2012ck}
K.~Miyamoto and K.~Nakayama, ``{Cosmological and astrophysical constraints on
  superconducting cosmic strings},''
  \href{http://dx.doi.org/10.1088/1475-7516/2013/07/012}{{\em JCAP} {\bfseries
  07} (2013) 012}, \href{http://arxiv.org/abs/1212.6687}{{\ttfamily
  arXiv:1212.6687 [astro-ph.CO]}}.

\bibitem{Hindmarsh:2022awe}
M.~Hindmarsh and J.~Kume, ``{Multi-messenger constraints on Abelian-Higgs
  cosmic string networks},'' \href{http://arxiv.org/abs/2210.06178}{{\ttfamily
  arXiv:2210.06178 [astro-ph.CO]}}.

\bibitem{Vincent:1997cx}
G.~Vincent, N.~D. Antunes, and M.~Hindmarsh, ``{Numerical simulations of string
  networks in the Abelian Higgs model},''
  \href{http://dx.doi.org/10.1103/PhysRevLett.80.2277}{{\em Phys. Rev. Lett.}
  {\bfseries 80} (1998) 2277--2280},
  \href{http://arxiv.org/abs/hep-ph/9708427}{{\ttfamily arXiv:hep-ph/9708427}}.

\bibitem{Moore:2001px}
J.~N. Moore, E.~P.~S. Shellard, and C.~J. A.~P. Martins, ``{On the evolution of
  Abelian-Higgs string networks},''
  \href{http://dx.doi.org/10.1103/PhysRevD.65.023503}{{\em Phys. Rev. D}
  {\bfseries 65} (2002) 023503},
  \href{http://arxiv.org/abs/hep-ph/0107171}{{\ttfamily arXiv:hep-ph/0107171}}.

\bibitem{Bevis:2006mj}
N.~Bevis, M.~Hindmarsh, M.~Kunz, and J.~Urrestilla, ``{CMB power spectrum
  contribution from cosmic strings using field-evolution simulations of the
  Abelian Higgs model},''
  \href{http://dx.doi.org/10.1103/PhysRevD.75.065015}{{\em Phys. Rev. D}
  {\bfseries 75} (2007) 065015},
  \href{http://arxiv.org/abs/astro-ph/0605018}{{\ttfamily
  arXiv:astro-ph/0605018}}.

\bibitem{Hindmarsh:2008dw}
M.~Hindmarsh, S.~Stuckey, and N.~Bevis, ``{Abelian Higgs Cosmic Strings: Small
  Scale Structure and Loops},''
  \href{http://dx.doi.org/10.1103/PhysRevD.79.123504}{{\em Phys. Rev. D}
  {\bfseries 79} (2009) 123504},
  \href{http://arxiv.org/abs/0812.1929}{{\ttfamily arXiv:0812.1929 [hep-th]}}.

\bibitem{Dufaux:2010cf}
J.-F. Dufaux, D.~G. Figueroa, and J.~Garcia-Bellido, ``{Gravitational Waves
  from Abelian Gauge Fields and Cosmic Strings at Preheating},''
  \href{http://dx.doi.org/10.1103/PhysRevD.82.083518}{{\em Phys. Rev. D}
  {\bfseries 82} (2010) 083518},
  \href{http://arxiv.org/abs/1006.0217}{{\ttfamily arXiv:1006.0217
  [astro-ph.CO]}}.

\bibitem{Hiramatsu:2013tga}
T.~Hiramatsu, Y.~Sendouda, K.~Takahashi, D.~Yamauchi, and C.-M. Yoo, ``{Type-I
  cosmic string network},''
  \href{http://dx.doi.org/10.1103/PhysRevD.88.085021}{{\em Phys. Rev. D}
  {\bfseries 88} no.~8, (2013) 085021},
  \href{http://arxiv.org/abs/1307.0308}{{\ttfamily arXiv:1307.0308
  [astro-ph.CO]}}.

\bibitem{Daverio:2015nva}
D.~Daverio, M.~Hindmarsh, M.~Kunz, J.~Lizarraga, and J.~Urrestilla,
  ``{Energy-momentum correlations for Abelian Higgs cosmic strings},''
  \href{http://dx.doi.org/10.1103/PhysRevD.95.049903}{{\em Phys. Rev. D}
  {\bfseries 93} no.~8, (2016) 085014},
  \href{http://arxiv.org/abs/1510.05006}{{\ttfamily arXiv:1510.05006
  [astro-ph.CO]}}. [Erratum: Phys.Rev.D 95, 049903 (2017)].

\bibitem{Correia:2018gew}
J.~R. C. C.~C. Correia and C.~J. A.~P. Martins, ``{Abelian-Higgs Cosmic String
  Evolution with CUDA},''
  \href{http://dx.doi.org/10.1016/j.ascom.2020.100388}{{\em Astron. Comput.}
  {\bfseries 32} (2020) 100388},
  \href{http://arxiv.org/abs/1809.00995}{{\ttfamily arXiv:1809.00995
  [physics.comp-ph]}}.

\bibitem{Correia:2020yqg}
J.~R. C. C.~C. Correia and C.~J. A.~P. Martins, ``{Abelian\textendash{}Higgs
  cosmic string evolution with multiple GPUs},''
  \href{http://dx.doi.org/10.1016/j.ascom.2020.100438}{{\em Astron. Comput.}
  {\bfseries 34} (2021) 100438},
  \href{http://arxiv.org/abs/2005.14454}{{\ttfamily arXiv:2005.14454
  [physics.comp-ph]}}.

\bibitem{Blanco-Pillado:2023sap}
J.~J. Blanco-Pillado, D.~Jim\'enez-Aguilar, J.~Lizarraga, A.~Lopez-Eiguren,
  K.~D. Olum, A.~Urio, and J.~Urrestilla, ``{Nambu-Goto Dynamics of Field
  Theory Cosmic String Loops},''
  \href{http://arxiv.org/abs/2302.03717}{{\ttfamily arXiv:2302.03717
  [hep-th]}}.

\bibitem{Kajantie:1998bg}
K.~Kajantie, M.~Karjalainen, M.~Laine, J.~Peisa, and A.~Rajantie,
  ``{Thermodynamics of gauge invariant U(1) vortices from lattice Monte Carlo
  simulations},'' \href{http://dx.doi.org/10.1016/S0370-2693(98)00440-7}{{\em
  Phys. Lett. B} {\bfseries 428} (1998) 334--341},
  \href{http://arxiv.org/abs/hep-ph/9803367}{{\ttfamily arXiv:hep-ph/9803367}}.

\bibitem{Blanco-Pillado:2011egf}
J.~J. Blanco-Pillado, K.~D. Olum, and B.~Shlaer, ``{Large parallel cosmic
  string simulations: New results on loop production},''
  \href{http://dx.doi.org/10.1103/PhysRevD.83.083514}{{\em Phys. Rev. D}
  {\bfseries 83} (2011) 083514},
  \href{http://arxiv.org/abs/1101.5173}{{\ttfamily arXiv:1101.5173
  [astro-ph.CO]}}.

\bibitem{Blanco-Pillado:2013qja}
J.~J. Blanco-Pillado, K.~D. Olum, and B.~Shlaer, ``{The number of cosmic string
  loops},'' \href{http://dx.doi.org/10.1103/PhysRevD.89.023512}{{\em Phys. Rev.
  D} {\bfseries 89} no.~2, (2014) 023512},
  \href{http://arxiv.org/abs/1309.6637}{{\ttfamily arXiv:1309.6637
  [astro-ph.CO]}}.

\bibitem{Blanco-Pillado:2017rnf}
J.~J. Blanco-Pillado, K.~D. Olum, and X.~Siemens, ``{New limits on cosmic
  strings from gravitational wave observation},''
  \href{http://dx.doi.org/10.1016/j.physletb.2018.01.050}{{\em Phys. Lett. B}
  {\bfseries 778} (2018) 392--396},
  \href{http://arxiv.org/abs/1709.02434}{{\ttfamily arXiv:1709.02434
  [astro-ph.CO]}}.

\bibitem{Kawasaki:2010yi}
M.~Kawasaki, K.~Miyamoto, and K.~Nakayama, ``{Gravitational waves from kinks on
  infinite cosmic strings},''
  \href{http://dx.doi.org/10.1103/PhysRevD.81.103523}{{\em Phys. Rev. D}
  {\bfseries 81} (2010) 103523},
  \href{http://arxiv.org/abs/1002.0652}{{\ttfamily arXiv:1002.0652
  [astro-ph.CO]}}.

\bibitem{Matsui:2016xnp}
Y.~Matsui, K.~Horiguchi, D.~Nitta, and S.~Kuroyanagi, ``{Improved calculation
  of the gravitational wave spectrum from kinks on infinite cosmic strings},''
  \href{http://dx.doi.org/10.1088/1475-7516/2016/11/005}{{\em JCAP} {\bfseries
  11} (2016) 005}, \href{http://arxiv.org/abs/1605.08768}{{\ttfamily
  arXiv:1605.08768 [astro-ph.CO]}}.

\bibitem{Matsui:2019obe}
Y.~Matsui and S.~Kuroyanagi, ``{Gravitational wave background from kink-kink
  collisions on infinite cosmic strings},''
  \href{http://dx.doi.org/10.1103/PhysRevD.100.123515}{{\em Phys. Rev. D}
  {\bfseries 100} no.~12, (2019) 123515},
  \href{http://arxiv.org/abs/1902.09120}{{\ttfamily arXiv:1902.09120
  [astro-ph.CO]}}.

\bibitem{Saikawa:2018rcs}
K.~Saikawa and S.~Shirai, ``{Primordial gravitational waves, precisely: The
  role of thermodynamics in the Standard Model},''
  \href{http://dx.doi.org/10.1088/1475-7516/2018/05/035}{{\em JCAP} {\bfseries
  05} (2018) 035}, \href{http://arxiv.org/abs/1803.01038}{{\ttfamily
  arXiv:1803.01038 [hep-ph]}}.

\bibitem{LIGOScientific:2017ikf}
{\bfseries LIGO Scientific, Virgo} Collaboration, B.~P. Abbott {\em et~al.},
  ``{Constraints on cosmic strings using data from the first Advanced LIGO
  observing run},'' \href{http://dx.doi.org/10.1103/PhysRevD.97.102002}{{\em
  Phys. Rev. D} {\bfseries 97} no.~10, (2018) 102002},
  \href{http://arxiv.org/abs/1712.01168}{{\ttfamily arXiv:1712.01168 [gr-qc]}}.

\bibitem{LIGOScientific:2014pky}
{\bfseries LIGO Scientific} Collaboration, J.~Aasi {\em et~al.}, ``{Advanced
  LIGO},'' \href{http://dx.doi.org/10.1088/0264-9381/32/7/074001}{{\em Class.
  Quant. Grav.} {\bfseries 32} (2015) 074001},
  \href{http://arxiv.org/abs/1411.4547}{{\ttfamily arXiv:1411.4547 [gr-qc]}}.

\bibitem{VIRGO:2014yos}
{\bfseries VIRGO} Collaboration, F.~Acernese {\em et~al.}, ``{Advanced Virgo: a
  second-generation interferometric gravitational wave detector},''
  \href{http://dx.doi.org/10.1088/0264-9381/32/2/024001}{{\em Class. Quant.
  Grav.} {\bfseries 32} no.~2, (2015) 024001},
  \href{http://arxiv.org/abs/1408.3978}{{\ttfamily arXiv:1408.3978 [gr-qc]}}.

\bibitem{Punturo:2010zz}
M.~Punturo {\em et~al.}, ``{The Einstein Telescope: A third-generation
  gravitational wave observatory},''
  \href{http://dx.doi.org/10.1088/0264-9381/27/19/194002}{{\em Class. Quant.
  Grav.} {\bfseries 27} (2010) 194002}.

\bibitem{Bartolo:2016ami}
N.~Bartolo {\em et~al.}, ``{Science with the space-based interferometer LISA.
  IV: Probing inflation with gravitational waves},''
  \href{http://dx.doi.org/10.1088/1475-7516/2016/12/026}{{\em JCAP} {\bfseries
  12} (2016) 026}, \href{http://arxiv.org/abs/1610.06481}{{\ttfamily
  arXiv:1610.06481 [astro-ph.CO]}}.

\bibitem{Yagi:2011wg}
K.~Yagi and N.~Seto, ``{Detector configuration of DECIGO/BBO and identification
  of cosmological neutron-star binaries},''
  \href{http://dx.doi.org/10.1103/PhysRevD.83.044011}{{\em Phys. Rev. D}
  {\bfseries 83} (2011) 044011},
  \href{http://arxiv.org/abs/1101.3940}{{\ttfamily arXiv:1101.3940
  [astro-ph.CO]}}. [Erratum: Phys.Rev.D 95, 109901 (2017)].

\bibitem{vanHaasteren:2011ni}
R.~van Haasteren {\em et~al.}, ``{Placing limits on the stochastic
  gravitational-wave background using European Pulsar Timing Array data},''
  \href{http://dx.doi.org/10.1111/j.1365-2966.2011.18613.x}{{\em Mon. Not. Roy.
  Astron. Soc.} {\bfseries 414} no.~4, (2011) 3117--3128},
  \href{http://arxiv.org/abs/1103.0576}{{\ttfamily arXiv:1103.0576
  [astro-ph.CO]}}. [Erratum: Mon.Not.Roy.Astron.Soc. 425, 1597 (2012)].

\bibitem{NANOGrav:2020bcs}
{\bfseries NANOGrav} Collaboration, Z.~Arzoumanian {\em et~al.}, ``{The
  NANOGrav 12.5 yr Data Set: Search for an Isotropic Stochastic
  Gravitational-wave Background},''
  \href{http://dx.doi.org/10.3847/2041-8213/abd401}{{\em Astrophys. J. Lett.}
  {\bfseries 905} no.~2, (2020) L34},
  \href{http://arxiv.org/abs/2009.04496}{{\ttfamily arXiv:2009.04496
  [astro-ph.HE]}}.

\bibitem{Janssen:2014dka}
G.~Janssen {\em et~al.}, ``{Gravitational wave astronomy with the SKA},''
  \href{http://dx.doi.org/10.22323/1.215.0037}{{\em PoS} {\bfseries AASKA14}
  (2015) 037}, \href{http://arxiv.org/abs/1501.00127}{{\ttfamily
  arXiv:1501.00127 [astro-ph.IM]}}.

\bibitem{Schmitz:2020syl}
K.~Schmitz, ``{New Sensitivity Curves for Gravitational-Wave Signals from
  Cosmological Phase Transitions},''
  \href{http://dx.doi.org/10.1007/JHEP01(2021)097}{{\em JHEP} {\bfseries 01}
  (2021) 097}, \href{http://arxiv.org/abs/2002.04615}{{\ttfamily
  arXiv:2002.04615 [hep-ph]}}.

\bibitem{Schmitz:2020}
K.~Schmitz, ``{New Sensitivity Curves for Gravitational-Wave Experiments
  (Version v1) [Data set]},''
  \href{http://dx.doi.org/https://doi.org/10.5281/zenodo.3689582}{{\em
  https://doi.org/10.5281/zenodo.3689582} (2020) }.

\end{thebibliography}\endgroup

\end{document}